\documentclass[journal]{vgtc}                     
\usepackage{soul}
\usepackage{colortbl} 
\usepackage{dblfloatfix}

\onlineid{0}

\vgtccategory{Research}

\vgtcpapertype{Journal}

\title{Large Language Model-assisted Speech and Pointing Benefits Multiple 3D Object Selection in Virtual Reality}

\author{%
  \authororcid{Junlong Chen}{0000-0002-7375-6525},
  \authororcid{Jens Grubert}{0000-0002-3858-2961}, and \authororcid{Per Ola Kristensson}{0000-0002-7139-871X}
}

\authorfooter{
  \item
        Junlong Chen and Per Ola Kristensson are with the University of Cambridge. E-mail: \{jc2375\,$|$\,pok21\}@cam.ac.uk\,.

  \item Jens Grubert is with Coburg University of Applied Sciences and Arts.
  	E-mail: jens.grubert@hs-coburg.de.
}

\abstract{%
  Selection of occluded objects is a challenging problem in virtual reality, even more so if multiple objects are involved.  With the advent of new artificial intelligence technologies, we explore the possibility of leveraging large language models to assist multi-object selection tasks in virtual reality via a multimodal speech and raycast interaction technique. We validate the findings in a comparative user study (n=24), where participants selected target objects in a virtual reality scene with different levels of scene perplexity. The performance metrics and user experience metrics are compared against a mini-map based occluded object selection technique that serves as the baseline. Results indicate that the introduced technique, \textsc{AssistVR}, outperforms the baseline technique when there are multiple target objects. Contrary to the common belief for speech interfaces, \textsc{AssistVR} was able to outperform the baseline even when the target objects were difficult to reference verbally. This work demonstrates the viability and interaction potential of an intelligent multimodal interactive system powered by large laguage models. Based on the results, we discuss the implications for design of future intelligent multimodal interactive systems in immersive environments.
}


\teaser{
  \centering
  \includegraphics[width=0.329\linewidth, alt={Low perplexity condition of AssistVR in the object selection user study.}]{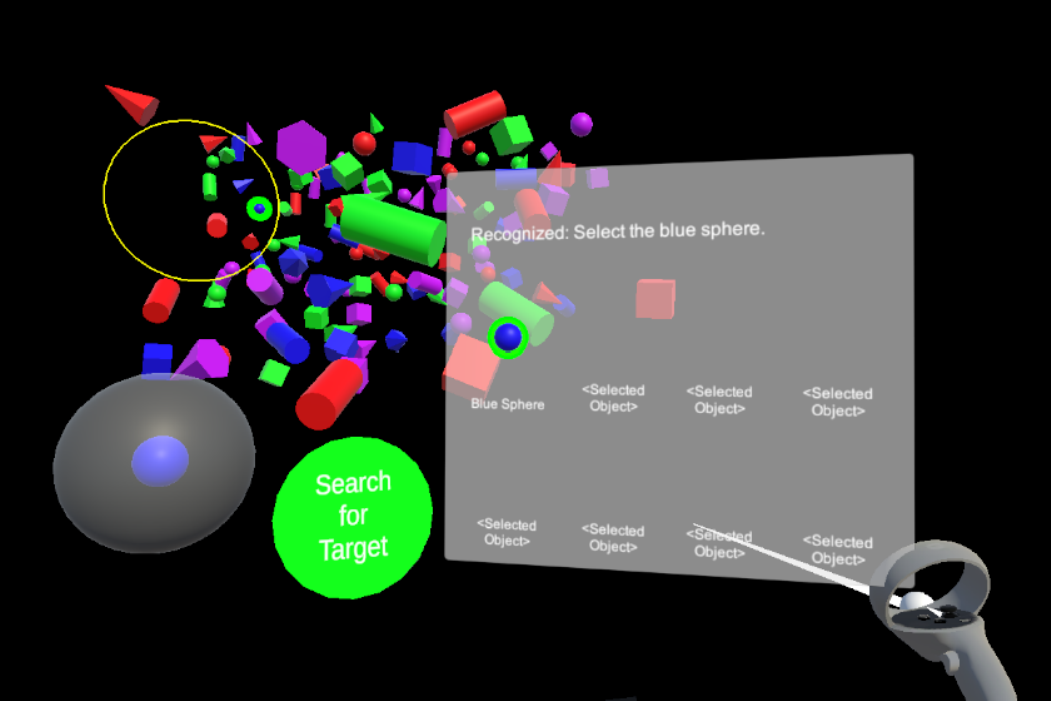}
  \includegraphics[width=0.329\linewidth, alt={Medium perplexity condition of AssistVR in the object selection user study.}]{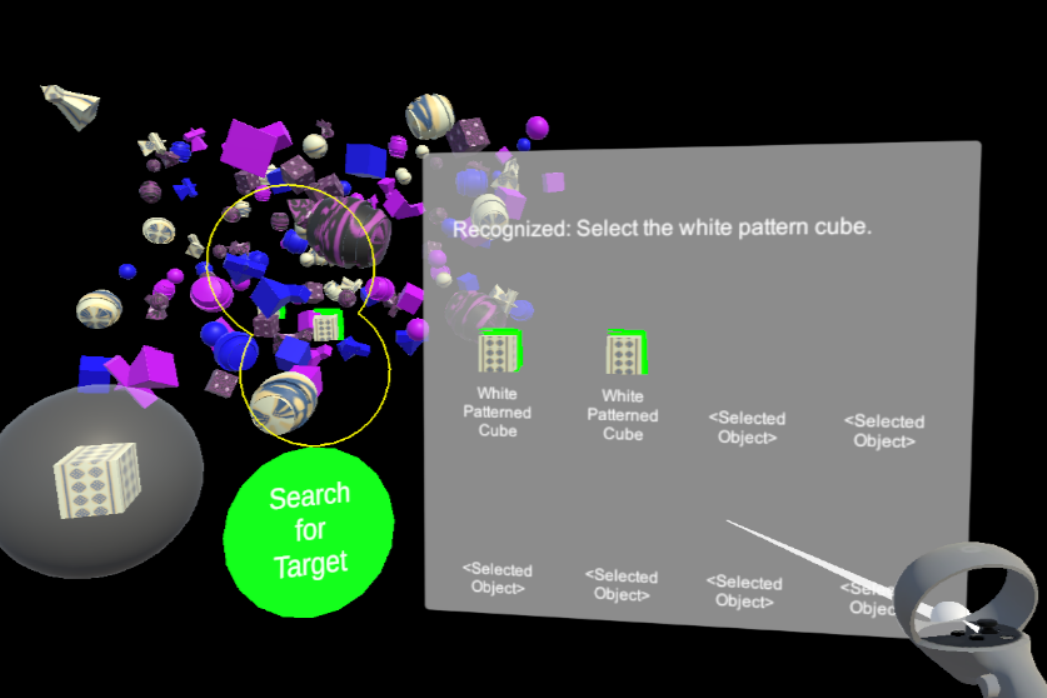}
  \includegraphics[width=0.329\linewidth, alt={High perplexity condition of AssistVR in the object selection user study.}]{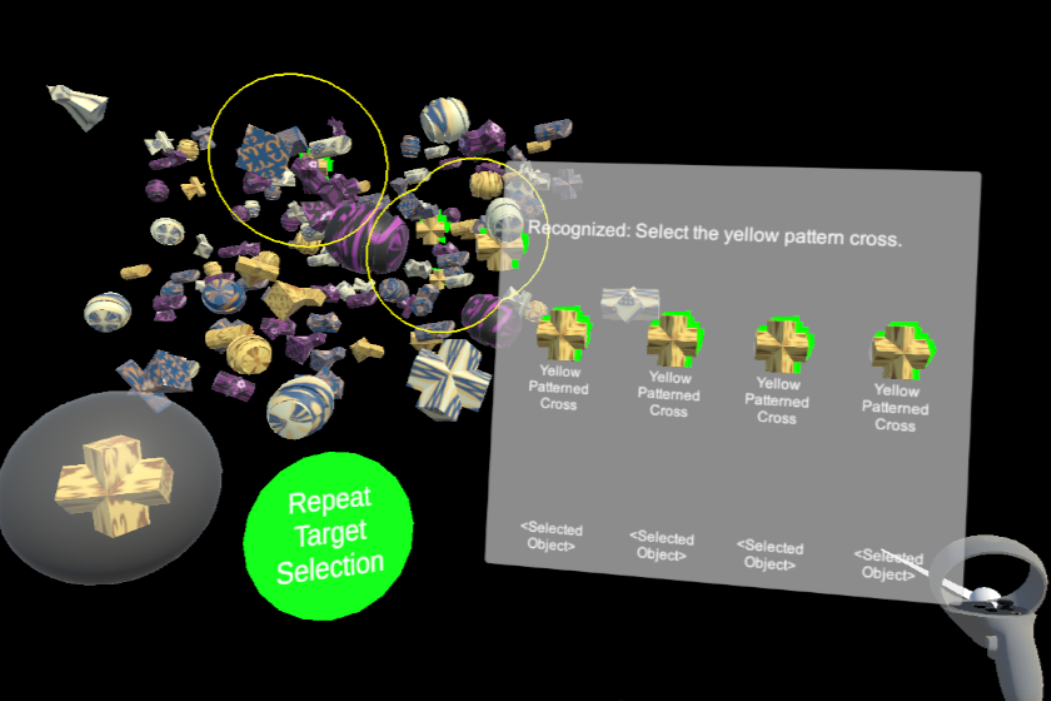}
  \caption{%
  	First-person views from the object selection user study with \textsc{AssistVR} under different scene perplexity conditions. Left: Low perplexity; Middle: Medium perplexity; Right: High perplexity.
  }
  \label{fig:teaser}
}

\graphicspath{{figs/}{figures/}{pictures/}{images/}{./}} 

\usepackage{tabu}                      
\usepackage{booktabs}                  
\usepackage{lipsum}                    
\usepackage{mwe}                       
\usepackage{balance}

\usepackage{mathptmx}                  

\begin{document}


\firstsection{Introduction}

\maketitle

Object selection in virtual reality is an important and widely researched task. Within this topic, there has been lots of interest on challenging sub-tasks, such as occluded object selection in virtual reality (VR). For occluded object selection alone, a myriad of tools have been developed to facilitate the task and improve user experience. These include a series of selection techniques developed by Yu et al.~\cite{yu2020fully}, as well as several works that followed which made improvements based on their original set of techniques to leverage gaze~\cite{yu2021gaze}, gestures~\cite{shi2023exploration}, or controllers~\cite{maslych2023toward} for occluded object selection and/or manipulation in virtual reality. Many more examples of works on occluded object selection in VR are reviewed in \Cref{sec:occluded}.

These works share many commonalities---they rely on the pointing metaphor and often focus on a single interaction modality to perform \textit{single} object selection. However, with the advent of generative artificial intelligence (GenAI) and large language models (LLMs), we can integrate intelligent conversational systems with traditional object selection techniques to develop a multimodal interaction technique for \textit{multi}-object selection under occluded conditions. While multimodal interaction techniques have been studied in interactive augmented and virtual environments for decades~\cite{bolt1980put, rakkolainen2021technologies}, LLMs promise to further enhance multimodal interaction, for example, by combining raycast and speech, as systems based on natural language can easily be integrated with LLMs. In an empirical study, we demonstrate that such a system yields certain advantages. Significant results indicate that by including speech-based interaction and AI models, we enhance the user's capability of completing challenging occluded object selection tasks in VR, especially under the condition when there are multiple objects to be selected, and when the target objects are easy to reference verbally. 

This paper bridges the research gap of efficient occluded \textit{multi}-object selection techniques in VR.
We leverage LLMs to complement traditional object selection techniques by proposing a conversation-based intelligent multimodal interactive system for the task of occluded object selection in VR.
We introduce an Advanced Speech Support and Interactive System for Virtual Reality (\textsc{AssistVR}), which combines a speech-based with a raycast-based interaction technique to perform 3D object selection tasks. 
We evaluate \textsc{AssistVR} on occluded object selection tasks in VR, and compare its performance and user experience ratings with a baseline technique. According to Yu et al.~\cite{yu2020fully}, several techniques exist for occluded object selection in VR. Among these different classes of techniques for occluded object selection, we selected a
minimap-based technique, \textsc{DiscPIM}~\cite{maslych2023toward}, as the baseline for the object selection task in this study.

The aim of the user study is to investigate the following research questions:

\begin{itemize}
    \item \textbf{RQ1:} Compared with the baseline technique \textsc{DiscPIM}~\cite{maslych2023toward}, how does \textsc{AssistVR} perform in terms of selection time and user experience when different numbers of target objects need to be selected?
    \item \textbf{RQ2:} Compared with the baseline technique \textsc{DiscPIM}~\cite{maslych2023toward}, how does \textsc{AssistVR} perform in terms of selection time and user experience under different scene perplexity conditions?
\end{itemize}



 To facilitate comparison with the DiscPIM baseline \cite{maslych2023toward},
 we recruited the same number of 24 participants in a within-subject user study and invited participants to complete a \textit{search} task and a \textit{repeat} task \cite{yu2020fully, maslych2023toward, petford2018pointing} for each combination of independent variables.
\textit{Search} and \textit{repeat} trial completion time serves as performance metrics, while ratings from questionnaires provide an indicator of user experience and task load.
Task completion performance and user experience results reveal that \textsc{AssistVR} required significantly less time to select two or more objects even when objects were difficult to reference verbally, and \textsc{AssistVR} was able to attain similar user experience ratings compared with the baseline technique.
Finally, the paper discusses the potential limitations of this work. It highlights areas for future work by providing recommendations to inform the design of intelligent speech-based interfaces for user interaction with immersive content.

In summary, this paper makes the following contributions:

\begin{itemize}
    \item \textbf{C1:} We present a multimodal interaction technique, \textsc{AssistVR}, which combines the raycast and speech modality to perform single and multi-object selection under occluded conditions in VR. 
    \item \textbf{C2:} We provide a proof-of-concept of integrating a customizable LLM to identify the intents and key entities in user speech input with a 3D interactive system for object selection in VR.
    \item \textbf{C3:} We compare the performance and user experience of our proposed method with a baseline occluded object selection method and distill findings from the user study to provide design recommendations for the design of interactive systems for immersive content based on language models. Specifically, we find that our technique \textsc{AssistVR} significantly outperforms a baseline technique in terms of task completion time for more than 2 targets (\textbf{30.5\%} faster for 2 targets and \textbf{86.4\%} faster for 4 targets).
\end{itemize}



\section{Related Work}
Our work is embedded in the areas of occluded object selection in VR, multimodal interaction techniques as well as work on large language models, which we will contextualize next.

\subsection{Occluded Object Selection in VR}\label{sec:occluded}

Object selection and manipulation are considered as fundamental interactions in VR \cite{mine1995virtual, poupyrev1999manipulating, vanacken2009multimodal, bergstrom2021evaluate}. They are often evaluated in testbed experiments \cite{bowman1999formalizing}, and many works have been dedicated to improve object selection \cite{li2020get, sidenmark2020outline} and manipulation \cite{zindulka2020performance} by proposing novel interaction techniques. Besides the virtual hand and raycast techniques, which are commonly used as interaction metaphors for selection and manipulation \cite{steed2005evaluating}, gestures \cite{li2020get} and eye gaze \cite{sidenmark2020outline} are also widely adopted in more recent literature for object selection.
Due to their high importance, these basic selection and manipulation tasks are widely studied in both virtual and augmented reality application scenarios \cite{li2022evaluation, pham2019pen}. Selection can be considered as the first step in the sequential process of referencing \cite{chastine2006framework}. Following Wei{\ss} et al. \cite{weiss2018user} and Sch{\"u}ssel et al. \cite{schussel2013influencing}, selection tasks provide a fundamental prerequisite for subsequent manipulation tasks, and findings on selection tasks will provide important implications for the design of interaction techniques when they are applied in other scenarios for other tasks.








In terms of object selection tasks, we are interested in how our technique performs under challenging selection scenarios, one of which being occluded object selection. Back in 2007, Vanacken et al. \cite{vanacken2007exploring} highlighted the research gap in dense and occluded object selection in 3D virtual environments and proposed the depth ray and the 3D bubble cursor to address this gap. Yu et al. \cite{yu2018target} studied the performance of different techniques in object selection under dense and occluded environments, and concluded that techniques enhanced with pointing facilitators can improve performance when there is no occlusion, which is not true when occlusions are present. Later in 2020, Yu et al. \cite{yu2020fully} developed a set of seven techniques (\textsc{Alpha Cursor}, \textsc{Flower Cone}, \textsc{Gravity Zone}, \textsc{Grid Wall}, \textsc{Lasso Grid}, \textsc{Magic Ball}, and \textsc{Smash Probe}) for fully-occluded target selection in VR, and studied how factors such as the number of occlusion layers, target depths, object densities, and estimated target locations could affect the performance of each technique. In the same year, Sidenmark et al. proposed \textsc{Outline Pursuits} \cite{sidenmark2020outline}, where candidate target objects within a cone are outlined with a stimulus moving along the outline. The occluded target object is then selected by matching the user's gaze movement with the movement of the stimulus on the outline. Subsequent works adopted ray-based metaphors such as \textsc{IntenSelect}  \cite{de2005intenselect} and \textsc{IntenSelect+} \cite{kruger2024intenselect+}, \textsc{LenSelect} \cite{weller2021lenselect}, \textsc{TouchRay} \cite{monteiro2023touchray}, \textsc{ClockRay} \cite{wu2023clockray}, redirected rays \cite{gabel2023redirecting}, and freehand pointing selection techniques \cite{shi2023exploration}. Additionally, ray metaphors can also be augmented by scene context \cite{wang2021scene}, minimap grabbing selections \cite{maslych2023toward}, and eye gaze \cite{chen2023gazeraycursor} for occluded object selection tasks.

Based on existing literature, we show that we can improve the above state-of-the-art object selection techniques by developing a speech and raycast multimodal technique for occluded multi-object selection tasks in VR and evaluating it against the state-of-the-art \textsc{DiscPIM} \cite{maslych2023toward} technique based on raycast and a minimap.


\subsection{Multimodal Interaction Techniques}

Back in 1980, Richard Bolt proposed ``Put-That-There" \cite{bolt1980put}, a voice and gesture multimodal interface for placing elements on a graphics display. The work became an example of how the speech modality could be applied in conjunction with other referencing modalities to provide an intuitive interaction experience for the user. Since then, a variety of multimodal systems emerged, and design principles for multimodal systems were distilled.

In 1999, Sharon Oviatt summarized ten myths in multimodal interaction \cite{oviatt1999ten}. For example, users like being able to interact with different modalities, but do not always do so. The work also pointed out that the speech and pointing multimodal interaction from ``Put-That-There" \cite{bolt1980put} and other deictic-point relations will have limited applications as it does not provide much useful functionality, and argues that speech should not be the primary input mode in multimodal systems. In 2004, Reeves et al. proposed a set of guidelines for the design of multimodal interfaces \cite{reeves2004guidelines} in terms of system requirements, multimodal input/output, adaptivity, consistency, feedback, and error prevention/handling.
During this period, Oviatt et al. \cite{oviatt2004we} also studied whether multimodal interfaces could help reduce user cognitive load. Their results indicated that users spontaneously shift to multimodal communication when task load increases, ultimately reaching a mix of unimodal and multimodal communication patterns - something we also observed during our study when participants where switching from single to multi-object selection tasks. Later in 2013, Sch{\"u}ssel et al. \cite{schussel2013influencing} studied influencing factors of multimodal interaction by conducting a Wizard of Oz study for object selection, and found a strong predominance in touch single modality input, and a rare occurrence in multimodal inputs for easy selection tasks. A review article by Matthew Turk \cite{turk2014multimodal} summarizes the above history of multimodal interaction research, and outlines challenges in sensing, recognition, usability, interaction, and privacy for multimodal HCI interface design. In more recent years, the introduction of virtual and augmented reality, as well as advanced sensing technologies and processing algorithms enable us to leverage gestures, facial expressions, eye gaze, audio, speech, and haptics to develop interactive systems which support many more modalities. Many review articles \cite{philippe2020multimodal, kim2021multimodal, rakkolainen2021technologies} summarize existing studies with multiple input and output modalities in virtual and augmented reality with a myriad of applications such as education, training, navigation, visualization, and monitoring.

\begin{figure*}[h!]
    \centering
    \includegraphics[width=0.9\linewidth]{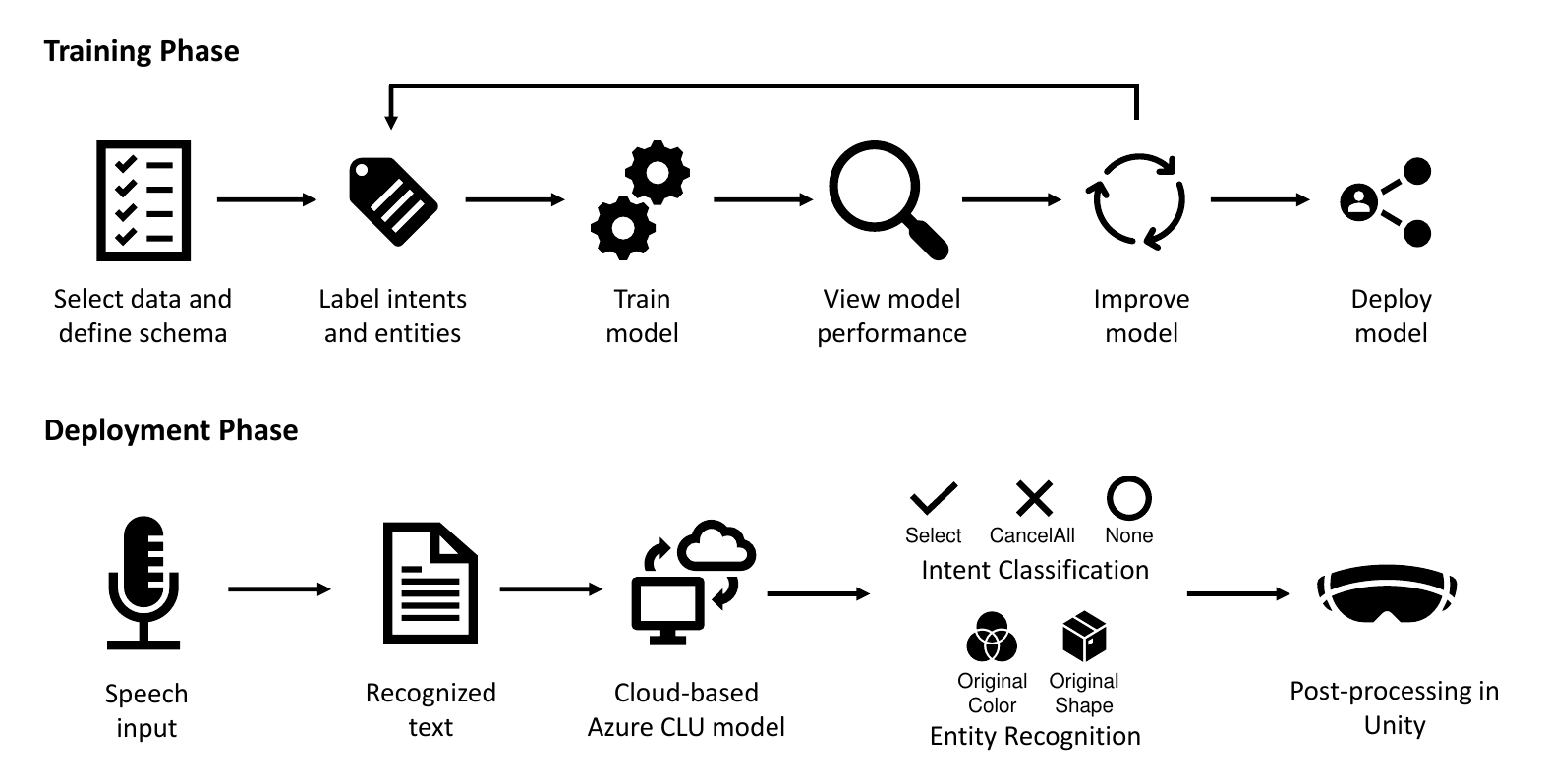}
    \caption{Workflow of the speech-based object selection component of \textsc{AssistVR} using Azure CLU. The workflow consists of the training phase (top row) and the deployment phase (bottom row), where a model is initially trained and then deployed to process user speech input in real-time.}
    \label{fig:workflow}
\end{figure*}

For the speech interaction modality specifically, Clark et al. \cite{clark2019state} conducted a thematic analysis to summarize nine primary research topics for HCI works with a focus in speech interfaces and identify gaps and challenges in speech interface research. Wei{\ss} et al. \cite{weiss2018user} provided a detailed comparison between 2D interfaces, 3D interfaces, and speech interfaces for VR. A quantitative user study consisting of tasks such as selection and simple manipulation is conducted, and the results indicate that while the speech interface held a significantly higher number of malfunctions than the 2D and 3D interface due to usage errors, parsing errors, and recognition errors, it is preferable if the interface needs to be easy to learn and involves lots of text entry. More recent works have different
focuses on speech interfaces. Lister et al. \cite{lister2020accessible} proposed design considerations for accessible conversational user interfaces (CUIs), Kim et al. \cite{kim2021designers} outlined challenges in designing voice user interfaces (VUIs) for natural human-agent conversation, while Guglielmi et al. \cite{guglielmi2024help} proposed VUI-UPSET, an approach for voice user interface testing.

This paper will focus on the interaction aspect of speech and raycast multimodal systems to study the performance and experience of users in VR object selection tasks, which will complement previous works on speech and multimodal interaction.

\subsection{Large Language Models}

Recent advances in transformer-based large language models have contributed substantially to the way 
natural language input can be processed. These systems typically fine-tune pre-trained language models on domain specific datasets to perform intent and entity recognition for specific applications or achieve language generation for general purposes.

In 2019, Devlin et al. introduced BERT \cite{devlin2018bert}, Bidirectional Encoder Representations from Transformers, which has achieved state-of-the-art performance on sentence-level and token-level tasks, paving the foundation for several intent classification and entity recognition systems. For example, Chen et al. \cite{chen2019bert} adopted BERT for natural language understanding (NLU) tasks such as joint intent classification and slot filling tasks. Jiang et al. \cite{jiang2021knowledge} leveraged BERT models to perform intent classification tasks based on speech input. BERT models are also widely adapted in named entity recognition tasks \cite{souza2019portuguese, chang2021chinese}. In the examples above, BERT models offer several advantages in customizability and interpretability. Developers are able to fine-tune models on customized data, and the identified intents and entities provide clear intermediate steps in the natural language understanding process, which in turn gives developers and users better interpretability and more agency over the system.

In contrast to
transformer-based NLP models 
like BERT, general-purpose LLMs like GPT-3 \cite{brown2020language} are trained on a much larger amount of data. While this allows LLMs to generate human-like responses directly from input prompts, they bring about other problems such as
lack of transparency \cite{zhao2024explainability} and limited customizability \cite{chen2024large}.

Based on the above features of both types of LLMs, this paper does not adopt the state-of-the-art general-purpose LLMs to process user speech input as they are difficult to finetune and interpret, and may occasionally generate plausible but incorrect information. Instead, we adopt an off-the-shelf Azure Conversational Language Understanding (CLU) model\footnote{Source: \url{https://azure.microsoft.com/en-us/products/ai-services/conversational-language-understanding}.}, a
customizable LLM to process natural language input from the user, as the model is easy to train and evaluate, while providing high accuracy and greater control in intent and entity recognition tasks. The structured output from the model also allows us to seamlessly integrate the CLU model with the interactive system in the 3D scene.

In this paper, we fill the gap in the existing literature by applying LLM techniques in a new domain - namely occluded object selection in VR. We contribute to the research community's understanding of object selection tasks in VR through the use of speech-based AI technologies.



\section{AssistVR Design Concept and Implementation}


The proposed object selection technique, \textsc{AssistVR}, is a multimodal selection method which consists of a speech-based selection technique using the Microsoft Azure CLU Application Programming Interface (API), together with a raycast-based selection technique. 
In the remainder of this section, we will introduce the implementation of the speech and raycast object selection modalities of \textsc{AssistVR} in further detail.

\subsection{Speech-based Selection}

The speech-based object selection technique constitutes the main component of \textsc{AssistVR}. We used the workflow in \Cref{fig:workflow}, separated into a training and a deployment phase, to implement it.

In the training phase, the first step is to select data and define \textit{intents} and \textit{entities}. Here, intent refers to the user's intent for each \textit{utterance}. Each utterance in the training and test set is labeled with one intent only. Within each utterance, there might be several words or phrases which are of interest. Words or phrases that share similar characteristics can be grouped together as one entity, and each utterance can contain one, several, or zero entities. As the task focuses on object selection, we focused on the following three intents: `Select', `CancelAll', and `None'. Entities used with the `Select' intent include `Original Color' and `Original Shape', while the `CancelAll' intent and the `None' intent do not include any entities. 

The second step is to create the training set by suggesting typical utterances for each of the above three intents and label all entities that appear in each utterance. For example, an utterance such as: `Select the purple cube.' would belong to the `Select' intent, with `purple' labeled as `Original Color' and `cube' labeled as `Original Shape'. The entire dataset consists of 51 utterances for the `Select' intent, and 6 utterances for the `CancelAll' intent. The complete set of utterances together with the labeled intents and entities for the training and test set can be found as a JSON file\footnote{The JSON file can be imported as a Conversational Language Understanding project to Azure to recover the project.} in the `Supplemental Materials' folder in the linked GitHub repository on OSF. Among all utterances for the `Select' intent, there are some variations in sentence structure, and the command does not follow a fixed sentence pattern as in traditional speech command interfaces. Subsequently, the model is trained, and several rounds of improvement are made to the custom utterance dataset and their intent and entity labels, such that the model attains a better accuracy in classifying intents and recognizing entities. Finally, the model is deployed on the Azure cloud server.


In the deployment phase, the user presses a button on the controller and speaks to the system via the VR headset. The Azure text-to-speech service then transcribes audio input into text. Requests are posted by Unity via the Azure API to upload the recognized user speech input to the model. The model then outputs the prediction in JSON format, which contains information on the most likely intent, recognized entities, as well as their confidence scores.

Based on this model output data, a Unity C\# post-processing script parses all selectable objects and retrieves the object material name. For commands with intent `Select', If the user command contains only the `Original Color' entity or the `Original Shape' entity, all objects whose material contains the color or shape property are selected. If the user command contains both the `Original Color' entity and the `Original Shape' entity, all objects whose material contains the color and shape property are selected. If the command intent is `CancelAll', all objects are deselected. If the command intent is `None', no action is performed. 

\begin{figure}[h!]
    \centering
    \includegraphics[width=\linewidth]{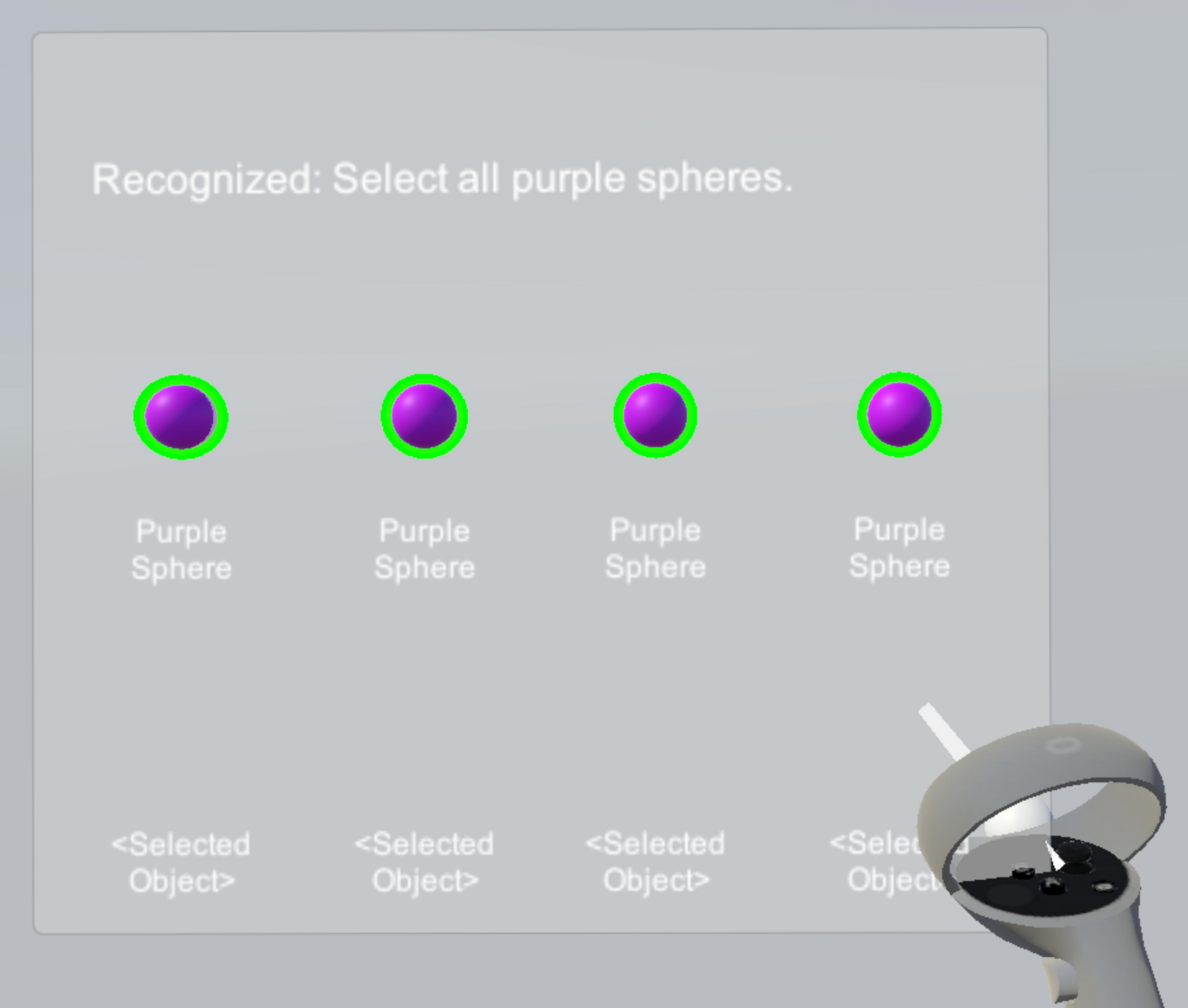}
    \caption{Draggable Panel of AssistVR which displays the recognized speech, together with a list of the names and 3D previews of all selected objects in the current scene. In this example, the user states `Select all purple spheres', and all four purple spheres in the scene are selected.}
    \label{fig:panel}
\end{figure}

Throughout the object selection process, all selected objects are displayed on a draggable panel (\Cref{fig:panel}) and are highlighted in green, both on the panel and in the 3D scene, to provide visual feedback to the user. When objects are deselected, they are removed from the panel. 
In addition to selected objects, the draggable panel also displays the recognized speech of the user. If the speech is not recognized, this information is also conveyed to the user via the panel.

\subsection{Raycast-based Selection}

As speech-based interfaces have the limitation of being prone to errors~\cite{weiss2018user, suhm2001multimodal} and usability tends to improve if speech-based interfaces are used in conjunction with traditional interfaces~\cite{bolt1980put, suhm2001multimodal}, we complement speech-based selection with a raycast technique to enable users to make fine-grained and precise selections/deselections.

Users are able to select/deselect objects hit by the ray which is cast from the right controller by pressing the trigger button. Selected objects are highlighted in green, while deselected objects are highlighted in red. When the ray moves away from a deselected object, the red highlight is removed. For selected objects, the green highlight is preserved even when the ray is directed away. As with objects selected using speech, objects selected using raycast also appear on the draggable panel with their names and green outlines.



\begin{figure*}[t!]
    \centering
    \includegraphics[width=0.329\textwidth]{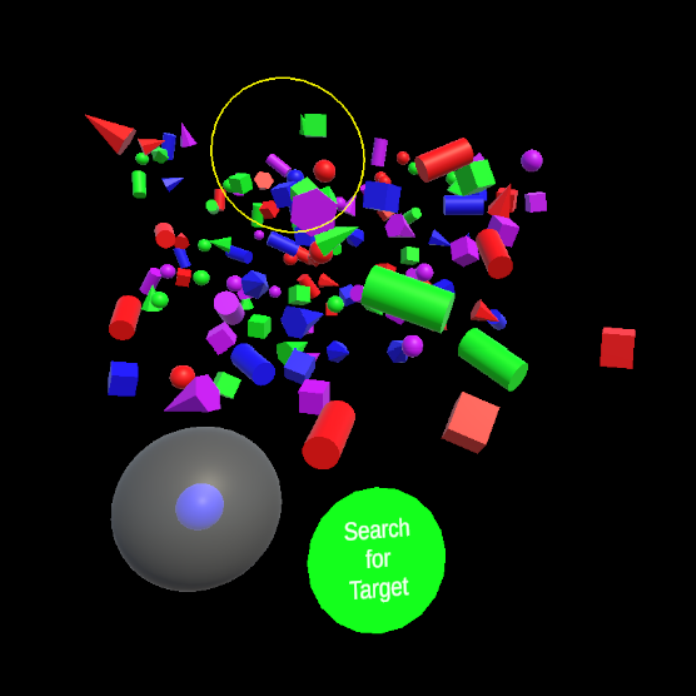}
    \includegraphics[width=0.329\textwidth]{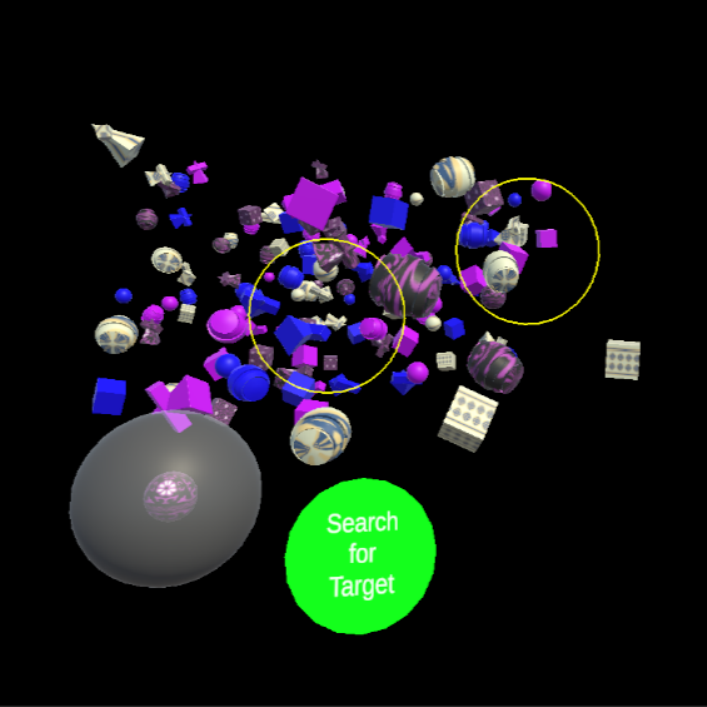}
    \includegraphics[width=0.329\textwidth]{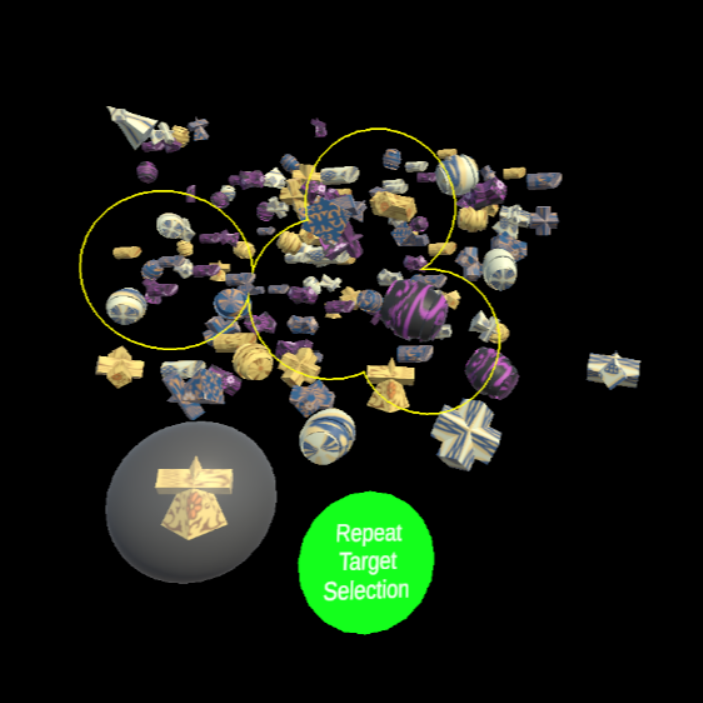}
    \caption{Illustration of the concept of scene perplexity applied in the object selection user study. From left to right: Low/Medium/High Scene Perplexity. The left image shows the user's view after starting the \textit{search} trial under the \textsc{1Target} condition. The middle image shows the view of the \textit{search} trial under the \textsc{2Targets} condition. The right image shows the view of the \textit{repeat} trial under the \textsc{4Targets} condition.}
    \label{fig:perplexity}
\end{figure*}

\section{Study Design}

The user study aims to evaluate the usability of our proposed speech and raycast multimodal 3D object selection technique (\textsc{AssistVR}) under different scene perplexity conditions and different numbers of target objects. Here, ``scene perplexity" refers to whether or not the object category and object property are known to the user. A detailed explanation is provided in Section \ref{sec:expt-design}. Our method is evaluated against a state-of-the-art occluded object selection method, \textsc{DiscPIM} \cite{maslych2023toward}, which serves as the baseline. This section introduces the design of the user study, which involves selection of different numbers of target objects under different scene perplexity conditions.
To facilitate comparison with the baseline technique \textsc{DiscPIM} \cite{maslych2023toward}, several elements of the experiment setup (such as the red/green button in front of the user, the yellow search region, the black background, and the search space dimensions) follow the setup proposed by Maslych et al. \cite{maslych2023toward}.

\subsection{Design}\label{sec:expt-design}


We adopt a within-subjects design to evaluate the performance of \textsc{AssistVR} and \textsc{DiscPIM}~\cite{maslych2023toward} in three levels of scene perplexity conditions (\textsc{Low}, \textsc{Medium}, and \textsc{High}), as well as three levels of target objects (\textsc{1Target}, \textsc{2Target}, and \textsc{4Target}). The scene perplexity is reflected by the number of object categories and object properties which can easily be referenced verbally. Examples of the \textsc{Low}, \textsc{Medium}, and \textsc{High} scene perplexity conditions can be found in \Cref{fig:perplexity}. In the user study, there are eight different object categories (of which four were easily recognizable by users and four were more difficult) and eight different textures to reflect different object properties (four were easily recognizable and four were more difficult). There are considerable differences between the known and unknown object categories and textures---a study with 21 participants prior to the study provides statistical evidence for the categorization of known and unknown objects' shapes and colors. Details of the survey findings can be found in the online appendix. 
Before the study, the names of all objects and textures are briefed to the participant to simulate real-world use cases where users have access to object names when using a speech-based system.

The \textsc{Low} perplexity condition consists of four easily-identifiable object shapes (cube, sphere, cylinder, and pyramid) and four easily-identifiable colors (purple, blue, green, and red)
The \textsc{Medium} perplexity condition consists of two easily-identifiable object shapes (cube and sphere) and two object shapes which are more difficult to identify (barrel and pyramid cuboid), as well as two easily-identifiable colors (purple and blue) and two colors that are more difficult to identify (purple pattern and white pattern). The \textsc{High} perplexity condition consisted of four object shapes which are difficult to identify (barrel, cross, pyramid cuboid, and truncated cylinder) and four colors which are difficult to identify (purple pattern, white pattern, yellow pattern, and blue pattern). The order of the two techniques and the order of the perplexity conditions are both counterbalanced across all 24 participants. 

The object density of distractor objects remains the same throughout all experiments.
In each scene perplexity level, 120 distractor objects are scattered within the environment of 20 meters in depth, 10 meters in width, and 5 meters in height in front of the user. Including the varying number of 1, 2, or 4 target objects, the total number of objects is 121, 122, or 124, depending on the \textsc{NumTargets} condition. The object shape and color of all objects depends on the scene perplexity level. Within each perplexity level, the shape, color, position, and orientation of distractor objects remains the same for different target objects.

\subsection{Task}

The task involves using the speech and raycast multimodal technique (\textsc{AssistVR}), or the baseline mini-map occluded object selection technique \textsc{DiscPIM}~\cite{maslych2023toward} to select different numbers of target objects (1, 2, or 4 targets) among a set of selectable objects. 
At the beginning of each trial, the user directs both the left and right raycast at a red button in front of the user\footnote{This step ensures that the user's raycast direction remains the same at the beginning of all trials~\cite{maslych2023toward}}. Upon pressing both triggers simultaneously to start the trial, the red button becomes green, the timer starts, and a home object\footnote{The home object is identical to the target object, which also serves to inform the user which object to select.} encapsulated within a semi-transparent sphere appears to the left of the user. At the same time, the target object(s), a yellow search region, together with 120 distractor objects appear within the 10m$\times$5m$\times$20m search space in front of the user. 
The shape and color of the target object(s) are randomly drawn from the set of 4 object shapes and 4 colors determined by the scene perplexity condition. Next, the participant uses the object selection technique to select a certain number of targets specified by the \textsc{NumTargets} condition. If an error occurs, participants are allowed to deselect and select again.
Finally, they press Button ``B'' on the right controller to confirm the selection. This stops the timer for trial completion time, and participants move on to the next trial if the selection is correct, that is, iff all target objects are selected and all selected objects are targets.

The time difference between starting the trial and users making the confirmation is recorded as the trial completion time.
Following prior work\cite{maslych2023toward,yu2020fully}, users are asked to repeat the same selection under the exactly same conditions after completion of the first selection trial. In this paper, we refer to these two tasks as the \textit{search} trial and the \textit{repeat} trial.

In summary, the study for each participant consists of 18 combinations of independent variables (2 \textsc{Techniques} $\times$ 3 \textsc{Perplexities} * $\times$ \textsc{NumTargets}). Participants complete the first half of the study with one technique, then complete the first half of the questionnaire, before moving on to the second technique and the second half of the questionnaire. Within each technique, participants complete trials under all three scene perplexities. Within each \textsc{Perplexity} condition, participants complete all three \textsc{NumTargets} conditions. Within each combination of \textsc{Technique}, \textsc{Perplexity}, and \textsc{NumTargets}, the participant completes three sets of one \textit{search} trial followed by one \textit{repeat} trial. For the \textit{search} trial in each set, the target object is randomly drawn from the list of 4 object shapes and 4 colors determined by the \textsc{Perplexity} condition for these three sets.

\subsection{Hypotheses}


We expect that the multimodal method will achieve similar performance regardless of the number of target objects to select, as speech commands should require a similar amount of time. Comparing the performance of our technique with the \textsc{DiscPIM} baseline, we expect that for a small number of target objects, the baseline method could achieve a better performance, whereas for a larger number of target objects, \textsc{AssistVR} could perform better. We formulate the following hypotheses with respect to the \textsc{NumTargets} condition:

\begin{itemize}
    \item \textbf{H1:} Participants take less time to complete the \textit{search} and \textit{repeat} trials with \textsc{AssistVR} compared to using \textsc{DiscPIM} when there are many targets, but this may not hold true when there is only a limited number of targets.
    \item \textbf{H2:} For \textsc{AssistVR}, the \textit{search} and \textit{repeat} trial completion time do not vary significantly when there are different numbers of target objects.
\end{itemize}

Following prior work \cite{wachsmuth2000using,borghi2018abstract,zhang2020study}, we expect that the multimodal selection method will perform well under low scene perplexity conditions, but may not perform as well as conventional selection techniques when the user finds it difficult to name the object category and/or property under high scene perplexity conditions. Based on this assumption, we formulate our hypotheses with respect to the \textsc{Perplexity} conditions as follows:

\begin{itemize}
    \item \textbf{H3:} Participants complete the \textit{search} and \textit{repeat} trial in less time with \textsc{AssistVR} under the \textsc{Low Perplexity} condition compared to using \textsc{DiscPIM}, but this may not hold true for the \textsc{Medium} and \textsc{High Perplexity} conditions.
    \item \textbf{H4:} For \textsc{AssistVR}, the \textit{search} and \textit{repeat} trial completion time is different under different \textsc{Perplexity} conditions.
\end{itemize}


\subsection{Participants and Apparatus}

We recruited 24 participants (16 males and 8 females) aged between 18 and 33 ($M=24.3\pm4.46$). All participants were right-handed and had normal or corrected-to-normal vision. Around 50\% of participants were familiar with head-mounted VR. All participants understood and spoke English, with around 58\% native English speakers. 
None of the participants reported any known visual, auditory, or physical disability. 
During the experiment, participants wore an Oculus Quest 2 headset and held the left and right controllers. The headset was connected to a Windows 10 laptop PC (Intel i5-9300H CPU, 16GB memory, and GTX 1050 graphics card) via cable. Virtual scenes were implemented with Unity 3D (Version 2020.3.47f1) and publicly available online asset resources. 

\begin{table*}[t]
\centering 
\small
\begingroup
\setlength{\tabcolsep}{5pt}
\begin{tabular}{|>{\centering}m{1.2cm}||>{\centering}m{0.2cm}|>{\centering}m{0.5cm}|c|c|c||>{\centering}m{0.2cm}|>{\centering}m{0.5cm}|c|c|c||>{\centering}m{0.2cm}|}
\multicolumn{11}{c}{\normalsize\bfseries\textbf{Trial Completion Time}} \\
\hline 
\multicolumn{1}{|c||}{}&
\multicolumn{5}{c||}{Search Task}& \multicolumn{5}{c||}{Repeat Task}\\
\hline 
 & d$f_{1}$ & d$f_{2}$ & F & p &  $\eta^2_p$ & d$f_{1}$ & d$f_{2}$ & F & p &  $\eta^2_p$\\ 
\hline 
T & \cellcolor{lightgray}$\mathbf{1}$ & \cellcolor{lightgray}$\mathbf{23}$ & \cellcolor{lightgray}$\mathbf{28.765}$ & \cellcolor{lightgray}$\mathbf{<.001}$ & \cellcolor{lightgray}$\mathbf{.556}$ &  
\cellcolor{lightgray}$\mathbf{1}$ & \cellcolor{lightgray}$\mathbf{23}$ & \cellcolor{lightgray}$\mathbf{47.329}$ & \cellcolor{lightgray}$\mathbf{<.001}$ & \cellcolor{lightgray}$\mathbf{.673}$ \\
N & \cellcolor{lightgray}$\mathbf{2}$ & \cellcolor{lightgray}$\mathbf{46}$ & \cellcolor{lightgray}$\mathbf{82.348}$ & \cellcolor{lightgray}$\mathbf{<.001}$ & \cellcolor{lightgray}$\mathbf{.782}$ &  
\cellcolor{lightgray}$\mathbf{2}$ & \cellcolor{lightgray}$\mathbf{46}$ & \cellcolor{lightgray}$\mathbf{147.450}$ & \cellcolor{lightgray}$\mathbf{<.001}$ & \cellcolor{lightgray}$\mathbf{.865}$ \\
P & \cellcolor{lightgray}$\mathbf{2}$ & \cellcolor{lightgray}$\mathbf{46}$ & \cellcolor{lightgray}$\mathbf{24.676}$ & \cellcolor{lightgray}$\mathbf{<.001}$ & \cellcolor{lightgray}$\mathbf{.518}$&  
\cellcolor{lightgray}$\mathbf{2}$ & \cellcolor{lightgray}$\mathbf{46}$ & \cellcolor{lightgray}$\mathbf{23.239}$ & \cellcolor{lightgray}$\mathbf{<.001}$ & \cellcolor{lightgray}$\mathbf{.503}$ \\
T $\times$ N & \cellcolor{lightgray}$\mathbf{2}$ & \cellcolor{lightgray}$\mathbf{46}$ & \cellcolor{lightgray}$\mathbf{34.053}$ & \cellcolor{lightgray}$\mathbf{<.001}$ & \cellcolor{lightgray}$\mathbf{.597}$ &  
\cellcolor{lightgray}$\mathbf{2}$ & \cellcolor{lightgray}$\mathbf{46}$ & \cellcolor{lightgray}$\mathbf{40.135}$ & \cellcolor{lightgray}$\mathbf{<.001}$ & \cellcolor{lightgray}$\mathbf{.636}$ \\
T $\times$ P & $2$ & $46$ & $.109$ & $.897$ & $.005$ &  
$2$ & $46$ & $1.784$ & $.179$ & $.072$ \\
T $\times$ N $\times$ P & $4$ & $92$ & $.549$ & $.700$ & $.023$ &  
$4$ & $92$ & $.453$ & $.770$ & $.019$ \\
\hline 
\end{tabular}

\endgroup

\caption{RM-ANOVA results for the trial completion time of both the \textit{search} and \textit{repeat} task after applying the Aligned Rank Transform. Gray rows show significant findings. T = \textsc{Interaction Technique}, N = \textsc{Number of Targets}, P = \textsc{Scene Perplexity}.}
\label{tab:PerformanceResults}
\end{table*}

\subsection{Procedure}\label{sec:procedure}

Before the study, participants were asked to review an information sheet and sign a consent form. After the collection of basic demographic information, participants were then asked to familiarize themselves with the Oculus Quest 2 headset and controllers. They were then given instructions on how to adjust the headstraps to comfortably put on the headset and learned how to use the controllers to perform selection commands using the two techniques. Next, we provided an overview of the study procedure and the tasks they were asked to complete using images and verbal introduction, which included explaining that the experiment will consist of using two techniques to perform object selection, and each technique consisted of three perplexity conditions, where each combination of \textsc{Technique} and \textsc{Perplexity} consisted of three sets of a \textit{search} trial and a \textit{repeat} trial.

We then demonstrated the usage of the two techniques to each participant. Participants had the opportunity to ask questions as we demonstrated how to use both controllers to press the start button and how to perform selection tasks using the two techniques. After answering all questions, participants had the opportunity to practice completing the trials with the two techniques under the \textsc{Medium Perplexity} condition.

After familiarization, participants completed three sets of \textit{search} and \textit{repeat} trials for each perplexity level and each number of targets using one of the two techniques, before having a five-minute break and completing another three sets for each perplexity level and each number of targets with the other technique. Throughout the experiment, the sequence of different perplexities and techniques were counterbalanced for all participants. In each trial, participants completed a \textit{search} task followed by a \textit{repeat} task. Specifically, they pointed the left and right raycast at the `Start' button and pressed the left and right trigger simultaneously to start a 3-second countdown to start the \textit{search} task. After the countdown, the target object drawn from different combinations of the set of object categories and object textures were generated at random locations within the 3D search space in front of the user. A yellow region indicates the approximate location of the target object. Users either use \textsc{AssistVR}, a combination of speech and raycast techniques to select the target object, or use \textsc{DiscPIM}~\cite{maslych2023toward} to generate a mini-map with the left controller, then use the right controller grip button to select the object directly from the mini-map, or from a list of expanded objects along the mini-map circumference if objects overlap in the mini-map. If the selection is correct, the timer will stop and the scene will reset. Users will then need to trigger the `Start' button again to begin the \textit{repeat} task. If the selection is incorrect, the system will play a tone to prompt the user to try again, and the total number of attempts will be recorded. In the subsequent \textit{repeat} task, the procedure is the same, except that the object positions are exactly the same as in the \textit{search} task, which are no longer randomly generated. After the \textit{search} and \textit{repeat} tasks, a new target object is drawn and placed at a new position, but the distractor objects remain at the same position.

After completing all trials for each technique, the participants were asked to complete SUS~\cite{brooke1996sus} and NASA-TLX~\cite{hart1988development} surveys. Following prior work~\cite{yu2020fully}, perceived user experience was measured using the short version of User Experience Questionnaire (UEQ-S)~\cite{schrepp2017design} on a 7-point scale. The surveys were followed by a five-minute break before progressing to the next technique. After completing trials for both techniques, participants were asked to rank the techniques based on their overall preference and provide feedback and comments on the features they preferred and disliked. The study took around 1.5 hours in total, and participants were compensated with a £15 Amazon voucher for their time.

\section{Results}

Statistical significance tests on trial completion time were carried out using a repeated measures analysis of variance (RM-ANOVA) with Holm-Bonferroni adjustments for the post-hoc tests. Task load, system usability, and user experience ratings were analyzed with non-parametric Wilcoxon signed-rank tests.


\subsection{Trial Completion Time}

During the study, we recorded the trial completion time as an indicator for object selection performance for both \textit{search} and \textit{repeat} tasks under all combinations of \textsc{Technique}, \textsc{Perplexity}, and \textsc{NumTargets} conditions as a quantitative measure of user performance in the object selection task.
In total, 2592 data points were collected (24 participants $\times$ 2 \textsc{Techniques} $\times$ 3 \textsc{Perplexities} $\times$ 3 \textsc{NumTargets} $\times$ 2 tasks $\times$ 3 repetitions). In line with prior work~\cite{maslych2023toward}, we removed 32 outlier data points (1.23\%) where the trial completion time was more than 4 standard deviations away from the mean in each condition. We did not discard trials which took participants more than one attempt to complete.
As each participant is exposed to all conditions, a repeated-measures ANOVA (RM-ANOVA) test was conducted on both the \textit{search} and \textit{repeat} trial completion time data to determine whether significant differences existed in trial completion time across different conditions.


\begin{figure}[t!]
    \centering
    \includegraphics[width=\linewidth]{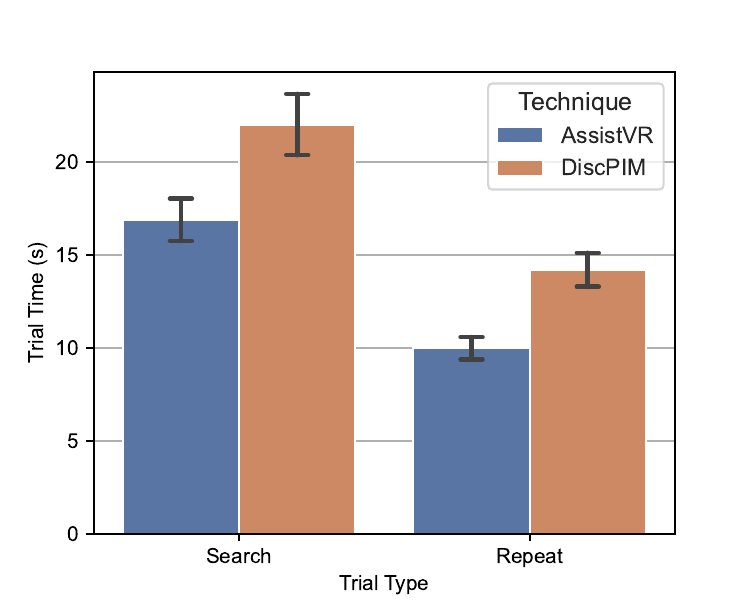}
    \caption{Bar plot of average trial completion time of \textsc{AssistVR} and \textsc{DiscPIM} \cite{maslych2023toward} in the \textit{search} and \textit{repeat} task. 95\% confidence intervals of the mean estimates are shown.}
    \label{fig:bar-technique}
\end{figure}

\begin{figure*}[t!]
    \centering
    \includegraphics[width=0.33\linewidth]{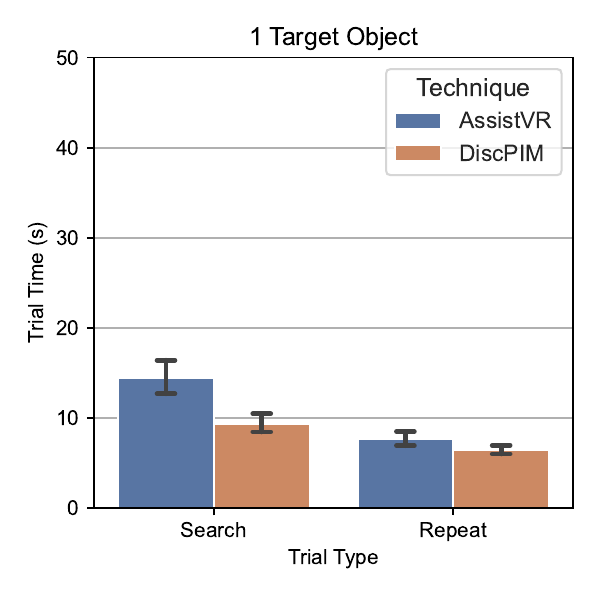}
    \includegraphics[width=0.33\linewidth]{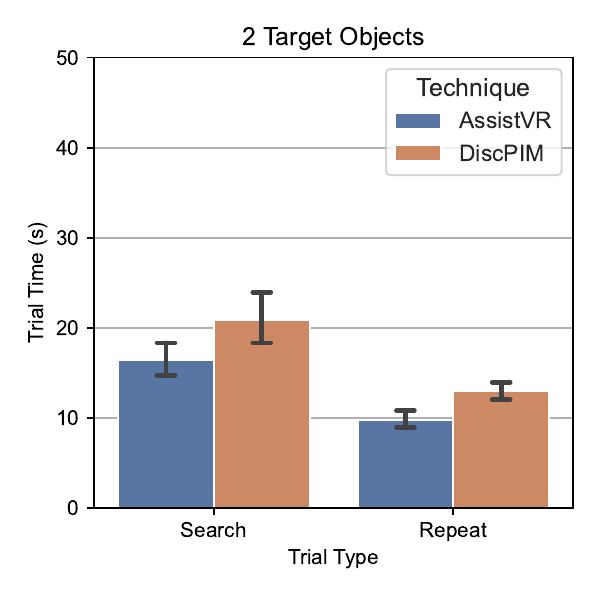}
    \includegraphics[width=0.33\linewidth]{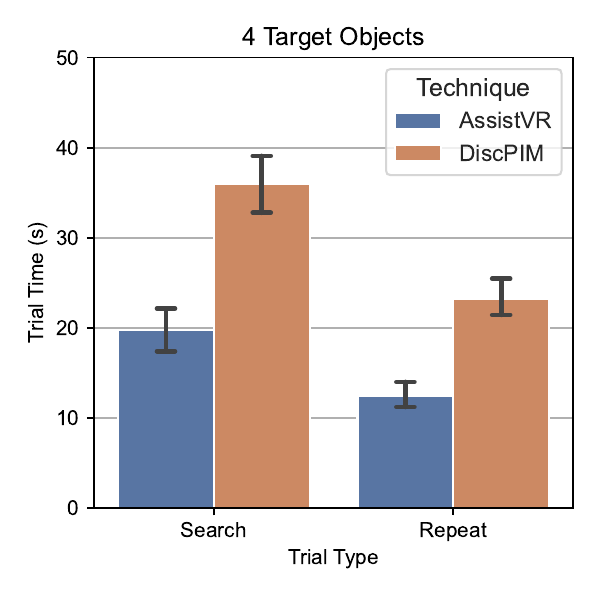}
    \caption{Trial completion time (seconds) for each technique across \textit{search} and \textit{repeat} trial types for each \textsc{NumTargets} condition, with 95\% confidence intervals of the mean estimate.
    }
    \label{fig:numTargets}
\end{figure*}

\begin{figure*}[t!]
    \centering
    \includegraphics[width=0.33\linewidth]{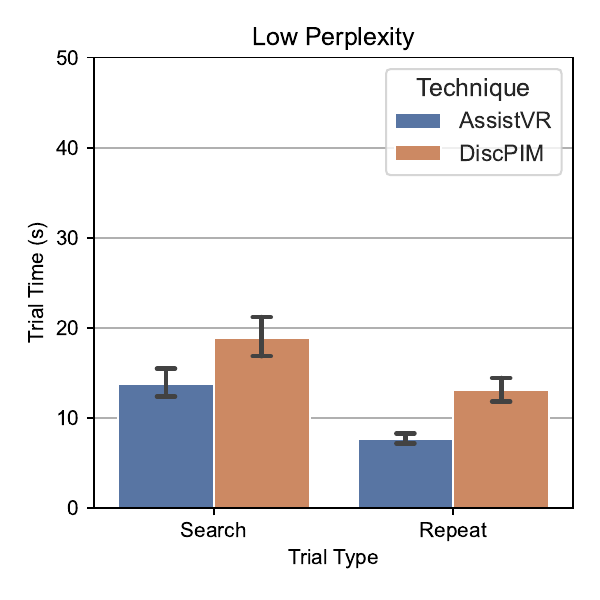}
    \includegraphics[width=0.33\linewidth]{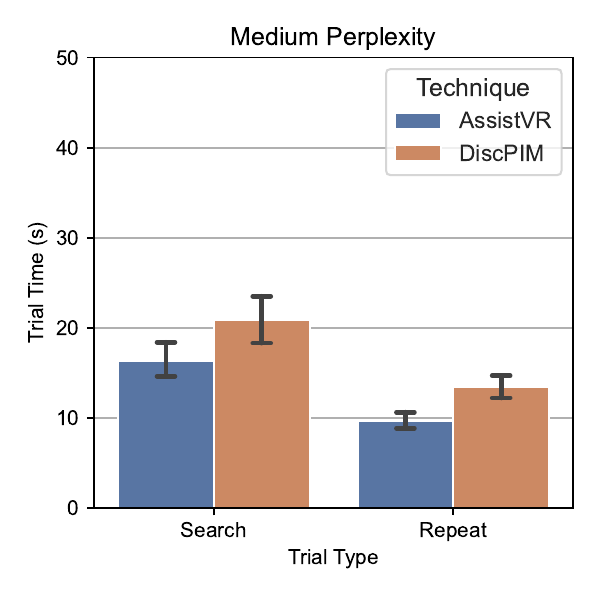}
    \includegraphics[width=0.33\linewidth]{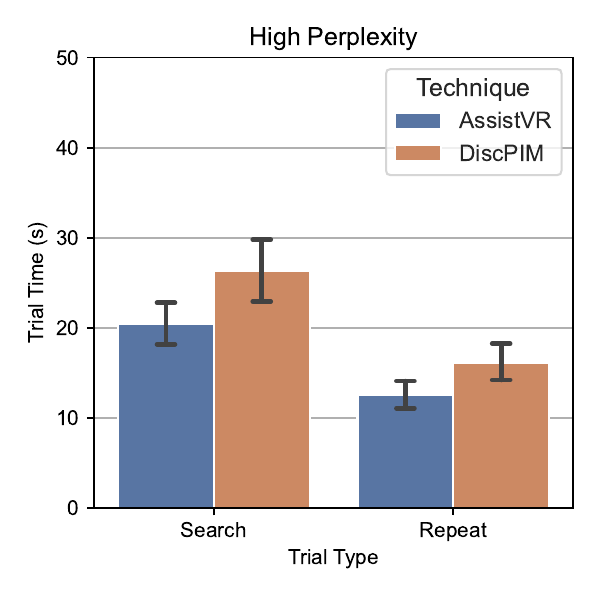}
    \caption{Trial completion time (seconds) for each technique across \textit{search} and \textit{repeat} trial types for each \textsc{Perplexity} condition, with 95\% confidence intervals of the mean estimate.
    }
    \label{fig:perplexity-result}
\end{figure*}



\Cref{tab:PerformanceResults} presents the RM-ANOVA results
on \textit{search} and \textit{repeat} trial completion time for the independent variables \textsc{Technique}, \textsc{NumTargets}, and \textsc{Perplexity}, together with interaction terms.

\subsubsection{Main Effect of \textsc{Technique}}

RM-ANOVA tests revealed a significant main effect of \textsc{Technique} on the \textit{search} ($F_{1,23}=28.765, \eta^2_p=.556, p < .001$) and \textit{repeat} ($F_{1,23}=47.329, \eta^2_p=.673, p < .001$) trial completion time.
\Cref{fig:bar-technique} presents the average \textit{search} and \textit{repeat} trial completion time for \textsc{AssistVR} and \textsc{DiscPIM} \cite{maslych2023toward} of all \textsc{Perplexity} and \textsc{NumTargets} conditions across all 24 participants. 
Post-hoc tests with Bonferroni adjustment suggested that participants took significantly less time ($p<.001$) to complete the \textit{search} task using \textsc{AssistVR} ($M=16.9, SD=9.79$) as opposed to using \textsc{DiscPIM} ($M=22.1, SD=16.3$). For the \textit{repeat} task, participants also took significantly less time ($p<.001$) with \textsc{AssistVR} ($M=10.1, SD=5.73$) compared with using \textsc{DiscPIM} ($M=14.3, SD=9.33$). Here, results for the main effect of \textsc{Technique} are averaged over the levels of \textsc{NumTargets} and \textsc{Perplexity}.

\subsubsection{Main Effect of \textsc{NumTargets}}

RM-ANOVA tests revealed a significant main effect of \textsc{NumTargets} on the \textit{search} ($F_{2,46}=82.348, \eta^2_p=.782, p < .001$) and \textit{repeat} ($F_{2,46}=147.450, \eta^2_p=.865, p < .001$) trial completion time.
Post-hoc tests with Bonferroni adjustment revealed that for the \textit{search} task, significant differences ($p<.001$) existed between all pairwise comparisons of the \textsc{1Target} ($M=11.9, SD=7.02$), \textsc{2Targets} ($M=18.7, SD=11.8$), and \textsc{4Targets} ($M=27.9, SD=15.7$) conditions. For the \textit{repeat} task, significant differences ($p<.001$) also existed in the trial completion time between all pairwise comparisons of the \textsc{1Target} ($M=7.09, SD=3.32$), \textsc{2Targets} ($M=11.5, SD=5.29$), and \textsc{4Targets} ($M=17.9, SD=9.72$) conditions.

\begin{table*}[t]
\centering 
\small
\begingroup
\setlength{\tabcolsep}{5pt}
\begin{tabular}{||>{\centering}m{1.2cm}|>{\centering}m{0.5cm}|>{\centering}m{0.5cm}|c||>{\centering}m{1cm}|>{\centering}m{0.5cm}|c|c||>{\centering}m{1cm}|>{\centering}m{0.5cm}|c|c||>{\centering}m{0.2cm}|}
\multicolumn{12}{c}{\normalsize\bfseries\textbf{Post-Experience Questionnaire Scores}} \\
\hline 
\multicolumn{4}{||c||}{NASA-TLX Scores}& \multicolumn{4}{c||}{SUS Scores}& \multicolumn{4}{c||}{UEQ-S Scores}\\
\hline 
 & $W$ & $p$ & $r$ & &  $W$ & $p$ & $r$  & & $W$ & $p$ & $r$ \\ 
\hline 
Overall & \cellcolor{lightgray}$\mathbf{59.5}$ & \cellcolor{lightgray}$\mathbf{<.05}$ & \cellcolor{lightgray}$\mathbf{.461}$ 
& Overall
& 157 & .853 & -0.038 
& Overall
& 106.5 & .346 & .197
\\
Mental & 76.0 & .103 & .348 
&
& &  &  & Pragmatic
& 176.5 & .457 & -.152 
\\
Physical & \cellcolor{lightgray}$\mathbf{55.5}$ & \cellcolor{lightgray}$\mathbf{<.05}$ & \cellcolor{lightgray}$\mathbf{.491}$ 
&
& &  &   
& Hedonic
& \cellcolor{lightgray}$\mathbf{64.5}$ & \cellcolor{lightgray}$\mathbf{<.05}$ & \cellcolor{lightgray}$\mathbf{.496}$ \\
Temporal & \cellcolor{lightgray}$\mathbf{2.0}$ & \cellcolor{lightgray}$\mathbf{<.05}$ & \cellcolor{lightgray}$\mathbf{.828}$ & 
& &  &  
&
& & & \\
Performance & 97.5 & .935 & -.019 &
& &  & 
&
&  & & \\
Effort & \cellcolor{lightgray}$\mathbf{37.5}$ & \cellcolor{lightgray}$\mathbf{<.05}$ & \cellcolor{lightgray}$\mathbf{.593}$ &
& &  & 
&
& & & \\
Frustration & 119.5 & .903 & -.027 & 
& &  & 
&
& & & \\
\hline 
\end{tabular}

\endgroup

\caption{Wilcoxon signed-rank test pairwise comparison results of NASA-TLX scores, SUS scores, and UEQ-S scores and their subcategories (if any) between both \textsc{Techniques}. Gray rows show significant findings. The Wilcoxon Statistic $W$, statistical significance $p$, and effect size $r$ are reported, where $r = z/\sqrt{N}$. A negative effect size indicates that the mean rating of \textsc{AssistVR} is higher than that of \textsc{DiscPIM}.}
\label{tab:QuestionnaireResults}
\end{table*}

\begin{figure*}[h!]
    \centering
    \includegraphics[height=2.58in]{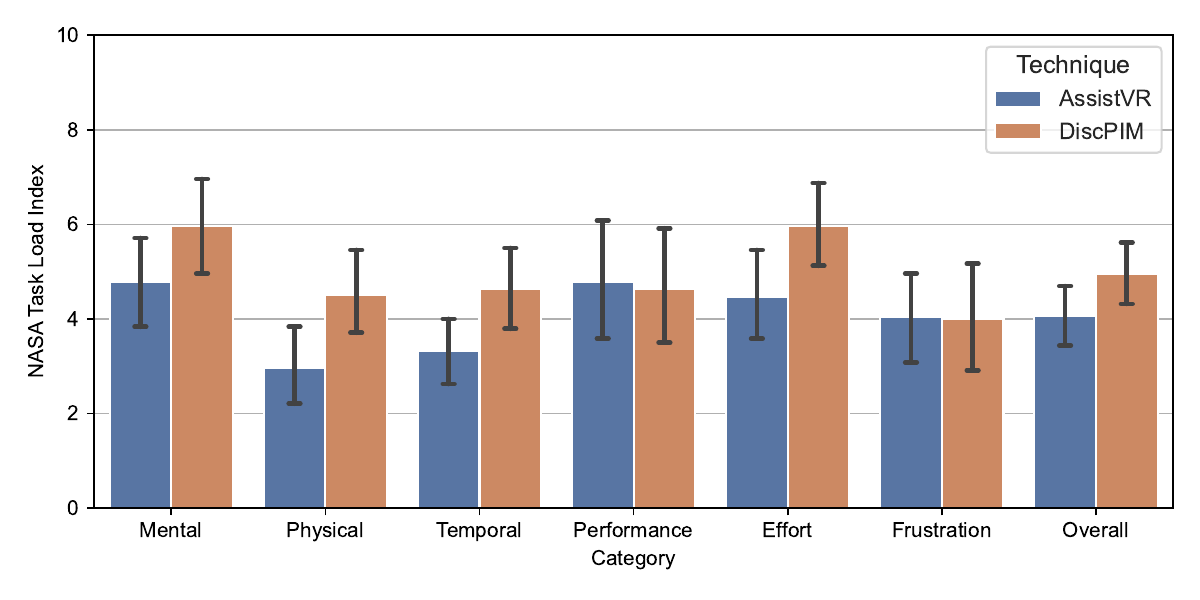}
    \includegraphics[height=2.58in]{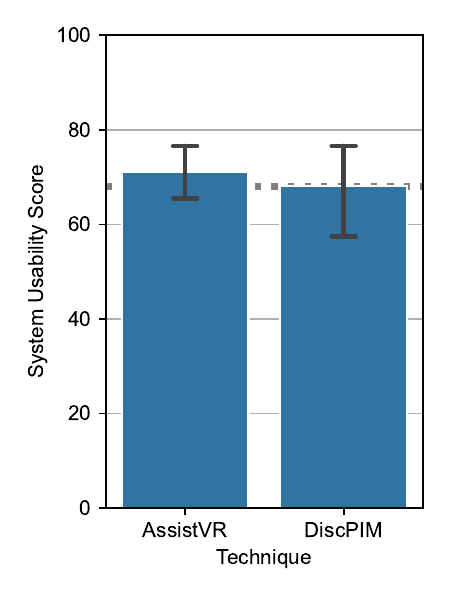}
    \caption{NASA-TLX ratings (left) and System Usability Scales (right) for the \textsc{AssistVR} and \textsc{DiscPIM} technique with 95\% confidence intervals. The grey dotted line in the SUS bar plot indicate the average SUS score of 68.}
    \label{fig:nasa-tlx}
\end{figure*}

\subsubsection{Main Effect of \textsc{Perplexity}}

RM-ANOVA tests revealed a significant main effect of \textsc{Perplexity} on the \textit{search} ($F_{2,46}=24.676, \eta^2_p=.518, p < .001$) and \textit{repeat} ($F_{2,46}=23.239, \eta^2_p=.503, p < .001$) trial completion time.
Post-hoc tests with Bonferroni adjustment also revealed that \textit{search} trial completion time was significantly different between the \textsc{High} perplexity condition ($M=23.4, SD=16.1$) and the \textsc{Low} perplexity condition ($M=16.5, SD=11.2$) ($p<.001$), between the \textsc{High} and \textsc{Medium} perplexity condition ($M=18.6, SD=12.3$) ($p<.05$), and between the \textsc{Low} and \textsc{Medium} perplexity condition ($p<.05$).
\textit{Repeat} trial completion time was significantly different between the \textsc{High} perplexity condition ($M=14.4, SD=9.58$) and the \textsc{Low} perplexity condition ($M=10.4, SD=6.75$) ($p<.001$), between the \textsc{High} and \textsc{Medium} perplexity condition ($M=11.7, SD=6.90$) ($p<.001$), but not between the \textsc{Low} and \textsc{Medium} perplexity condition ($p=.145$).

\subsubsection{Interaction Effect of \textsc{Technique} $\times$ \textsc{NumTargets}}


\Cref{fig:numTargets} shows the \textit{search} and \textit{repeat} trial completion time of different \textsc{Technique} and \textsc{NumTargets} combinations.

RM-ANOVA tests revealed a significant interaction effect of \textsc{Technique} $\times$ \textsc{NumTargets} on the \textit{search} ($F_{2,46}=34.053, \eta^2_p=.597, p < .001$) and \textit{repeat} ($F_{2,46}=40.135, \eta^2_p=.636, p < .001$) trial completion time.
For the \textsc{AssistVR} technique, post-hoc tests with Bonferroni adjustment did not reveal significant differences in \textit{search} trial completion time between the \textsc{1Target} and \textsc{4Targets} condition ($p=.088$), the \textsc{1Target} and \textsc{2Targets} condition ($p=1.0$), or the \textsc{2Targets} and \textsc{4Targets} condition ($p=.854$). Significant differences were found in \textit{repeat} trial completion time between the \textsc{1Target} and \textsc{2Targets} ($p<.05$) and \textsc{1Target} and \textsc{4Targets} ($p<.001$) conditions, but not between the \textsc{2Targets} and \textsc{4Targets} ($p=.069$) condition.


For the \textsc{DiscPIM} technique, post-hoc tests revealed that there is a significant difference in \textit{search} and \textit{repeat} trial completion time when the number of targets varied ($p<.001$ for all pairwise comparisons of \textsc{NumTargets} conditions in both \textit{search} and \textit{repeat} tasks).

Comparing the trial completion time between \textsc{AssistVR} and \textsc{DiscPIM} under different \textsc{NumTargets} conditions, post-hoc comparisons revealed that in the \textit{search} task, \textsc{DiscPIM} ($M=9.34, SD=4.62$) was significantly faster than \textsc{AssistVR} ($M=14.5, SD=8.03$) under the \textsc{1Target} condition ($p<.05$), but \textsc{AssistVR} ($M=19.8, SD=11.6$) was significantly faster than \textsc{DiscPIM} ($M=35.9, SD=15.1$) under the \textsc{4Targets} condition ($p<.001$). The difference in \textit{search} trial completion time between both techniques is not significant under the \textsc{2Target} condition ($p=.128$).
In the \textit{repeat} task, \textsc{AssistVR} ($M=9.96, SD=5.02$) was significantly faster than \textsc{DiscPIM} ($M=13.0, SD=5.13$) under the \textsc{2Targets} condition ($p<.05$). \textsc{AssistVR} ($M=12.5, SD=6.81$) was also significantly faster than \textsc{DiscPIM} ($M=23.3, SD=9.23$) under the \textsc{4Targets} condition ($p<.001$). The difference in \textit{repeat} trial completion time between both techniques is not significant under the \textsc{1Target} condition ($p=.603$).

\subsubsection{Interaction Effect of \textsc{Technique} $\times$ \textsc{Perplexity}}

\Cref{fig:perplexity-result} presents bar plots of the \textit{search} and \textit{repeat} trial completion time for different combinations of \textsc{Technique} and \textsc{Perplexity} conditions.
RM-ANOVA tests did not reveal a significant interaction effect of \textsc{Technique} $\times$ \textsc{Perplexity} on either the \textit{search} ($F_{2,46}=.109, \eta^2_p=.005, p=.897$) or \textit{repeat} ($F_{2,46}=1.784, \eta^2_p=.072, p=.179$) trial completion time. 

\subsubsection{Interaction Effect of \textsc{Technique} $\times$ \textsc{NumTargets} $\times$ \textsc{Perplexity}}

RM-ANOVA tests did not reveal a significant interaction effect of \textsc{Technique} $\times$ \textsc{NumTargets} $\times$ \textsc{Perplexity} on either the \textit{search} ($F_{4,92}=.549, \eta^2_p=.023, p=.700$) or \textit{repeat} ($F_{4,92}=.453, \eta^2_p=.019, p=.770$) trial completion time.

\subsection{Task Load}

\Cref{fig:nasa-tlx} (left) shows a bar plot of the NASA-TLX ratings (unweighted version) \cite{hart1988development} for each category as well as the overall load with 95\% confidence intervals of the mean score.
A Wilcoxon signed rank test revealed that the overall task load rating of \textsc{AssistVR} ($M=4.06, SD=1.71$) was significantly lower ($W=59.5, p<.05, r=.461$) than that of \textsc{DiscPIM} ($M=4.94, SD=1.72$). Within subcategory ratings, \textsc{Physical} load was found to be significantly lower ($W=55.5, p<.05, r=.491$) for \textsc{AssistVR} ($M=2.96, SD=2.12$) compared to \textsc{DiscPIM} ($M=4.50, SD=2.17$). The \textsc{Temporal} load of \textsc{AssistVR} ($M=3.33, SD=1.81$) was significantly lower ($W=2.0, p<.05, r=.828$) than \textsc{DiscPIM} ($M=4.62, SD=2.32$), and the \textsc{Effort} rating of \textsc{AssistVR} ($M=4.46, SD=2.47$) was also significantly lower ($W=37.5, p<.05, r=.593$) than \textsc{DiscPIM} ($M=5.96, SD=2.31$). Results are summarized in \Cref{tab:QuestionnaireResults}.

\subsection{System Usability}

\Cref{fig:nasa-tlx} (right) presents a bar plot of the system usability scale \cite{brooke1996sus} of the \textsc{AssistVR} and \textsc{DiscPIM} technique with 95\% confidence intervals of the mean estimate. 
A Wilcoxon signed rank test did not reveal any significant differences ($W=157, p=.853, r=-.038$) between the SUS ratings of \textsc{AssistVR} ($M=71.0, SD=14.4$) and \textsc{DiscPIM} ($M=68.0, SD=23.0$). Results are summarized in \Cref{tab:QuestionnaireResults}.

\subsection{User Experience}

\begin{figure}[h!]
    \centering
    \includegraphics[width=\linewidth]{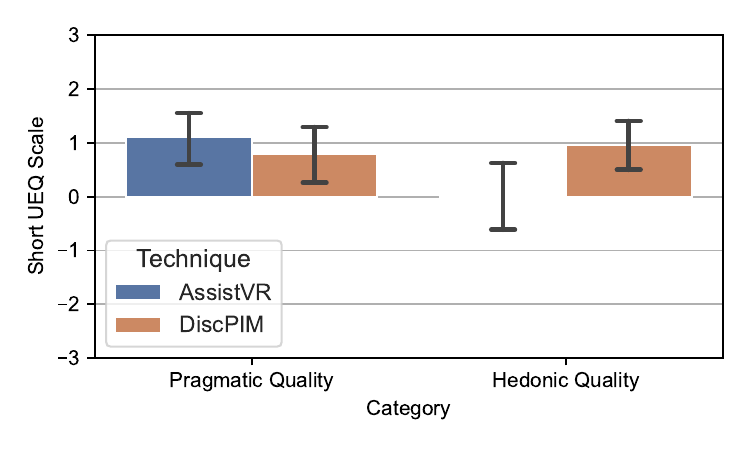}
    \caption{Bar plot of the Short UEQ Pragmatic and Hedonic Quality Scale for the \textsc{AssistVR} and \textsc{DiscPIM} technique with 95\% confidence intervals.
    }
    \label{fig:ueq-s}
\end{figure}

\Cref{fig:ueq-s} shows the results from the short version User Experience Questionnaire (UEQ-S) \cite{schrepp2017design},
where \textsc{DiscPIM} attains a higher average hedonic quality score
and \textsc{AssistVR} attains a higher average pragmatic quality score. 
Wilcoxon signed rank tests did not reveal a significant difference ($W=106.5, p=.346, r=.197$) in the overall UEQ-S score between \textsc{AssistVR} ($M=.547, SD=1.23$) and \textsc{DiscPIM} ($M=.880, SD=1.07$). For the subcategories of the UEQ-S ratings, significant differences ($W=64.5, p<.05, r=.496$) were found in the \textsc{Hedonic} quality between \textsc{AssistVR} ($M=-0.021, SD=1.61$) and \textsc{DiscPIM} ($M=.958, SD=1.16$), but not in the \textsc{Pragmatic} quality ($W=176.5, p=.457, r=-.152$) between \textsc{AssistVR} ($M=1.11, SD=1.21$) and \textsc{DiscPIM} ($M=.802, SD=1.37$). Results are summarized in \Cref{tab:QuestionnaireResults}.

\subsection{Overall Preference and Open Comments}

In the post-experience questionnaire, we asked participants about their overall preference among the two techniques and invited them to leave comments about features they liked/disliked. Among all 24 participants, 13 preferred the \textsc{AssistVR} technique, while 11 preferred the \textsc{DiscPIM} technique. 

For the \textsc{AssistVR} technique, participants liked the fact that it was easy to use (P7), efficient (P10, P20, P21, P22), and allowed `selecting multiple objects in one go' (P2, P14), and participants could select objects without knowing where the object is (P13, P14, P15, P23), or moving their hands to execute any action (P12). P19 also found that `the combination of speech and raycast stroke a nice balance', as raycast was more efficient for selecting one or two visible objects and voice selection helped to select multiple objects. Participants disliked the fact that it `did not support many commands' (P5). Further, speech recognition sometimes failed and the command was not executed correctly (P1, P2, P7, P11, P13, P21, P24). Consequently, the system `either doesn't select anything or selects wrong objects' (P19),  which led to frustration (P7) and loss of trust (P19). Specifically, P8, P12, P21, P22 commented that speech recognition was sensitive to accent, without the capability to auto-correct recognized speech based on the context (P22), which repeatedly led to errors. Some participants found it somewhat difficult to remember object names (P12, P13, P14, P19). Sometimes participants had to repeat several times before getting the speech command right (P5, P23). P20 also commented that the time it takes to speak and the slight delay in speech processing makes it sometimes faster to engage in manual selection as opposed to using speech. P24 also suggested that the \textsc{Deselect} function could deselect a specific object, rather than deselecting all objects.

For the \textsc{DiscPIM} technique, participants liked the fact that the design of the torch and mini-map felt `fun' (P14) and `intuitive' (P7, P12) and the mini-map freezes to provide direct visual feedback (P15, P16), which gives users a better sense of control and direct manipulation (P1, P2, P5, P19), and improves selection accuracy (P18). It `makes objects far away easier to see' (P23), allowing users to focus on a complicated region with many occluded objects (P10), without the need to know the object's name (P13). P4 also found \textsc{DiscPIM} to be `more engaging and less repetitive than speech recognition'. Regarding limitations, participants found \textsc{DiscPIM} to be `tedious' (P1), `annoying' (P2), `tricky and boring' (P5), and `slower' (P14) when there are many targets. P4 found the technique `tiring for the eyes' and `time consuming'. P19 suggested that the mini-map object expansion list could appear closer to the left hand or allow users to customize the position, which would otherwise require users to raise the head to look at it which is not very convenient. P20 and P21 commented that the benefits brought by \textsc{DiscPIM} is situational and it could slow down the search process when there is not much occlusion.

\section{Discussion}

This paper contributes to fill a gap in the literature addressing the  challenge of occluded \textit{multi}-object selection tasks in VR by developing and studying multimodal LLM-enabled interaction. Based on existing techniques leveraging ray-based metaphors~\cite{yu2020fully, de2005intenselect, kruger2024intenselect+, weller2021lenselect, monteiro2023touchray, wu2023clockray, gabel2023redirecting, maslych2023toward}, gestures~\cite{shi2023exploration}, and eye gaze~\cite{sidenmark2020outline, chen2023gazeraycursor} for object selection in VR, together with works studying speech interaction~\cite{weiss2018user, lister2020accessible, kim2021designers}, multimodal interaction~\cite{philippe2020multimodal, kim2021multimodal, rakkolainen2021technologies} in immersive technologies, as well as customizable purpose-built LLMs~\cite{chen2024large}, we have advanced the research community's understanding of the design of intelligent multimodal interactive systems for object selection in VR by proposing \textsc{AssistVR} and validating user performance and experience by comparing it against a baseline technique, \textsc{DiscPIM}, in an empirical user study. 

Key findings from the study show that users were able to select objects in VR faster with \textsc{AssistVR} when there were multiple objects, and the object perplexity (whether or not objects were difficult to reference verbally) did not compromise the high selection efficiency of \textsc{AssistVR} as long as participants had access to the names of all target objects. Comparing different conditions within \textsc{AssistVR}, we found that the number of targets and object perplexity both affected the selection completion time to a certain extent. Results indicate that the speech and raycast multimodal selection technique \textsc{AssistVR} posed a significantly lower overall load on users and provide a similar level of user experience quality compared with the baseline. 


For the independent variable \textsc{NumTargets}, 
while \textsc{DiscPIM} required significantly less \textit{search} time in the \textsc{1Target} condition, \textsc{AssistVR} required significantly less time in the \textsc{4Targets} condition in the \textit{search} trial, and significantly less time in the \textsc{2Targets} and \textsc{4Targets} conditions in the \textit{repeat} trial,
which provides evidence to support \textbf{H1}. 
No significant differences were found in the \textsc{AssistVR} \textit{search} trial completion time between all pairwise comparisons of \textsc{NumTargets} conditions. However, significant differences were found in the \textsc{AssistVR} repeat trial completion time between the \textsc{1Target} and \textsc{2Targets} as well as the \textsc{1Target} and \textsc{4Targets} condition. Therefore, we are unable to determine the validity of \textbf{H2} based on the quantitative data from the performance study.

In terms of the \textsc{Perplexity} condition,
performance results reveal 
that compared with using \textsc{DiscPIM}, participants required significantly less time to complete the \textit{repeat} task with \textsc{AssistVR} under
the \textsc{Low}, \textsc{Medium}, and \textsc{High} perplexity conditions. Therefore, we find that \textsc{AssistVR} outperforms the expectation listed in \textbf{H3}, given the assumption that participants know how to reference complex objects and have instant access to their names when they forget.
As results indicated a significant difference in \textit{search} completion time between the \textsc{Low} and \textsc{High} perplexity condition using \textsc{AssistVR}, as well as a significant difference in \textit{repeat} trial completion time between the \textsc{Low} and \textsc{High}, as well as \textsc{Medium} and \textsc{High} perplexity condition using \textsc{AssistVR}, we have evidence to support \textbf{H4}.
These performance results indicated that user performance with \textsc{AssistVR} was sensitive to the scene \textsc{Perplexity} condition. 



User experience questionnaires revealed that in terms of user experience, both techniques have advantages and disadvantages in different aspects. In terms of efficiency, participants reflected that \textsc{AssistVR} was efficient for multi-object selection, without the need to know the object's location or to search for it visually. However, the trade-off is that participants needed to memorize the names of the objects and reference it correctly. This posed many issues when speech recognition was inaccurate or when the speech processing algorithm failed to associate the speech command with an object of interest.
In terms of user agency and gamification, participants preferred \textsc{DiscPIM} as the interaction design felt fun and intuitive, and users had a better sense of control on object selection.
Based on the NASA-TLX survey results, \textsc{AssistVR} posed a significantly lower overall load on the participants compared with \textsc{DiscPIM}.
Based on SUS and UEQ-S data, users provided similar user experience ratings on both techniques.

from the above analysis, we distill the following design recommendations for \textsc{AssistVR} and similar speech-based multimodal interactive systems for immersive content:

     \textbf{DR1:} Our experiment results suggest that by incorporating the speech modality with another interaction modality that provides more precision, we are able to improve user performance in object referencing when there are multiple target objects. While it is a reasonable conjecture that speech-based interactions may fail when users do not know how to reference objects verbally, our results suggest that when users have easy access to the names of objects, multimodal interaction techniques with speech can outperform traditional referencing techniques even when object names are long and difficult to pronounce. 
     
     \textbf{DR2:} Our findings support prior research on multimodal speech-based interactive systems, which conclude that users do not necessarily use all modalities in a multimodal system, and speech-based systems can fail when speech recognition is not accurate~\cite{vertanen2009parakeet,kristensson2011asynchronous}. We propose that speech-based systems could improve speech recognition accuracy by leveraging contextual information in past conversations and support recognition of different accents. Post-processing pipelines of recognized intents and entities should be robust to all types of user inputs.
    
    \textbf{DR3:} Qualitative feedback from the user study suggests that speech-based multimodal interactive systems could ensure visibility of the mapping between speech commands and subsequent actions of the system to allow users to have agency~\cite{coyle2012did} over the system, and improve user experience by making speech interactions more engaging.

\subsection{Limitations and Future Work}

\textsc{AssistVR} is based on a natural language processing model, which poses requirements on the quality of custom utterance data. The quality of data input by the developer directly affects the performance of the model. In this study, as only three user intents are represented (\textsc{Select}, \textsc{CancelAll}, and \textsc{None}), the model is able to achieve a high accuracy on intent classification and entity recognition. However, for more complex tasks which involve more intents and entities, we expect that errors may exist in intent classification and entity recognition, which poses a limitation in generalizability.
 
Further, the nature of NLP models determine that the developer can only pre-define a limited number of intents and entities, and it is highly likely that when the system is deployed among a large number of users, these pre-defined intents and entities may not handle all user inputs properly. For example, in the selection task, the command `Deselect the red sphere' would be categorized under the intent `DeselectAll', even if the user only intends to deselect certain objects.
     
In addition, \textsc{AssistVR} had limited visibility, which resulted in a poor sense of user agency. Compared with \textsc{DiscPIM} where instant visual feedback was present, \textsc{AssistVR} only revealed a list of all selected objects. Users were unable to use the raycast method to parse through elements in the scene and inspect their properties in detail.
    
Finally, some participants reflected that while \textsc{AssistVR} could be efficient in certain scenarios, it was not as engaging as \textsc{DiscPIM} due to the fact that it only supported certain types of speech commands.

These limitations lead to the following avenues of future work. First, we see potential in extending the application of \textsc{AssistVR} to more complicated tasks, such as scene editing and manipulation, where users are required to reference objects, inspect associated information, and modify properties. The speech processing model would be required to support more intents and entities, which is more similar to real-world application scenarios.

Second, it would be useful to introduce an additional processing module to handle exceptions from the speech processing model. The additional module should be able to handle all speech commands, and execute subsequent actions directly or provide feedback to guide the user to supported commands.

Third, an additional component could be added to the raycast function of \textsc{AssistVR} to allow users to visually inspect the properties of various objects in the scene and provide instant feedback on object information.

 Fourth, \textsc{AssistVR} could be improved to support all sorts of speech commands, and respond in different tones and styles to engage the user.

\section{Conclusion}

In this paper, we propose the design of \textsc{AssistVR}, a multimodal interaction technique in VR based on speech and raycast. The proposed method allows users to perform object selection tasks based on raycast and controller button selection/deselection, which is accurate but less efficient, or using speech commands, which is efficient but does not provide users with the same sense of agency and transparency. We demonstrate the proof-of-concept of applying a speech recognition module and an off-the-shelf LLM, Azure CLU, to process user speech input and select virtual objects in a 3D scene. The user study results indicate that this method is able to outperform the baseline occluded object selection technique, \textsc{DiscPIM}, when there are multiple objects to select. The performance of \textsc{AssistVR} is not compromised even when objects are difficult to reference verbally. User experience ratings indicate that our proposed method is able to provide a lower level of perceived task load and a similar level of user experience as the baseline method. Findings from the user study suggest that we can further improve speech-based intelligent multimodal interactive systems by focusing on aspects such as user agency, system visibility, speech recognition and processing robustness, and system redundancy.

\section*{Supplemental Materials}
\label{sec:supplemental_materials}

Supplemental materials can be found on OSF at 
\url{https://osf.io/fh9ba/}
, which are released under a Creative Commons By Attribution 4.0 International license.
These include: (1) The pre-experiment survey and anonymized results, (2) Questionnaire for the user study on object selection, (3) Training and testing data for the Azure CLU system, and (4) The online appendix for this paper.

\acknowledgments{%
This work is supported by the China Scholarship Council (CSC) Cambridge Scholarship.
The authors would like to thank the authors of \textsc{DiscPIM} \cite{maslych2023toward} for open-sourcing their code to facilitate the study design and comparison against the baseline technique in this paper.
}

\bibliographystyle{abbrv-doi-hyperref}

\bibliography{main}
\balance


\vfill

\end{document}




\teaser{
    \centering
    \includegraphics[height=0.94in]{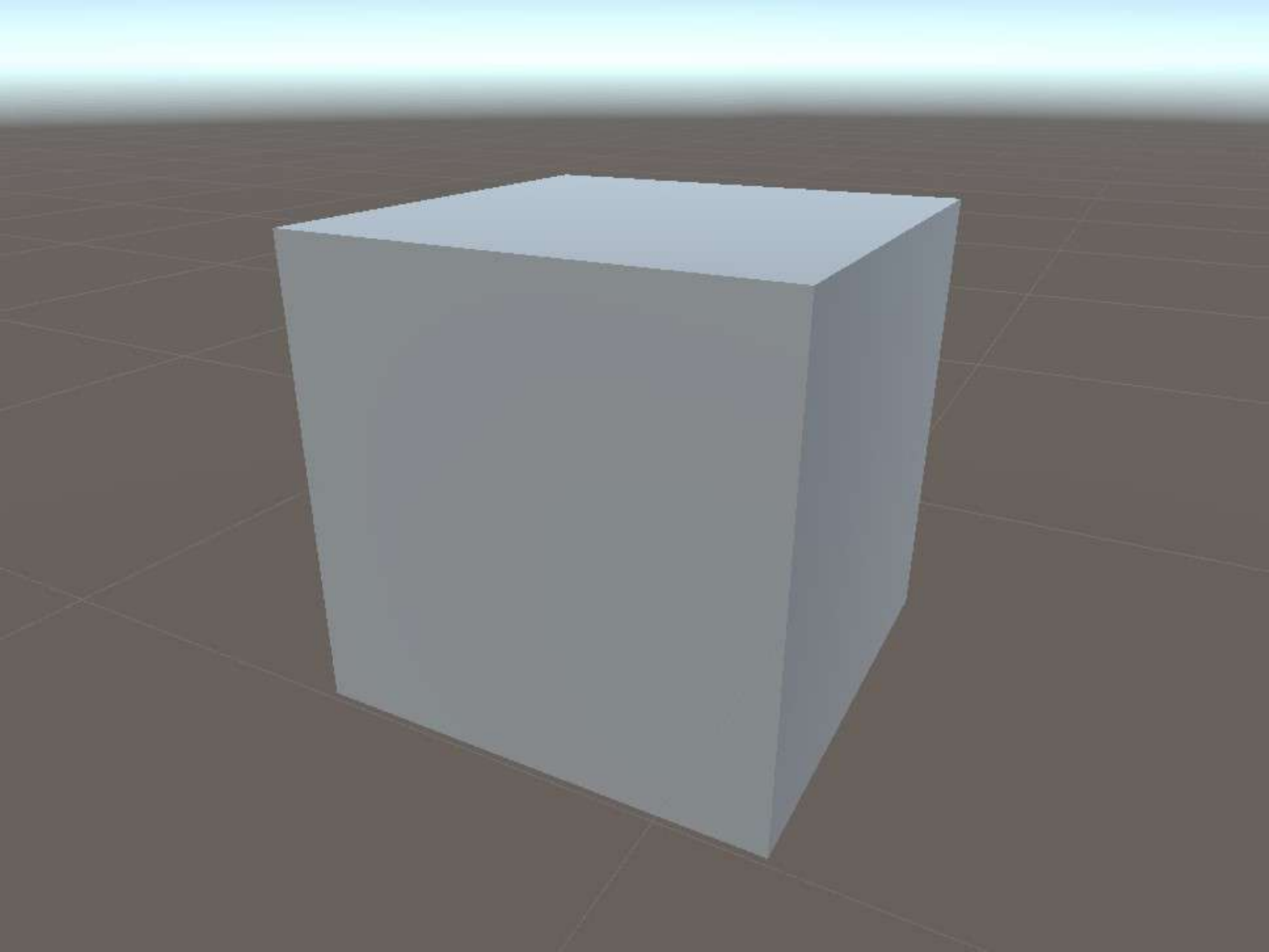}
    \includegraphics[height=0.94in]{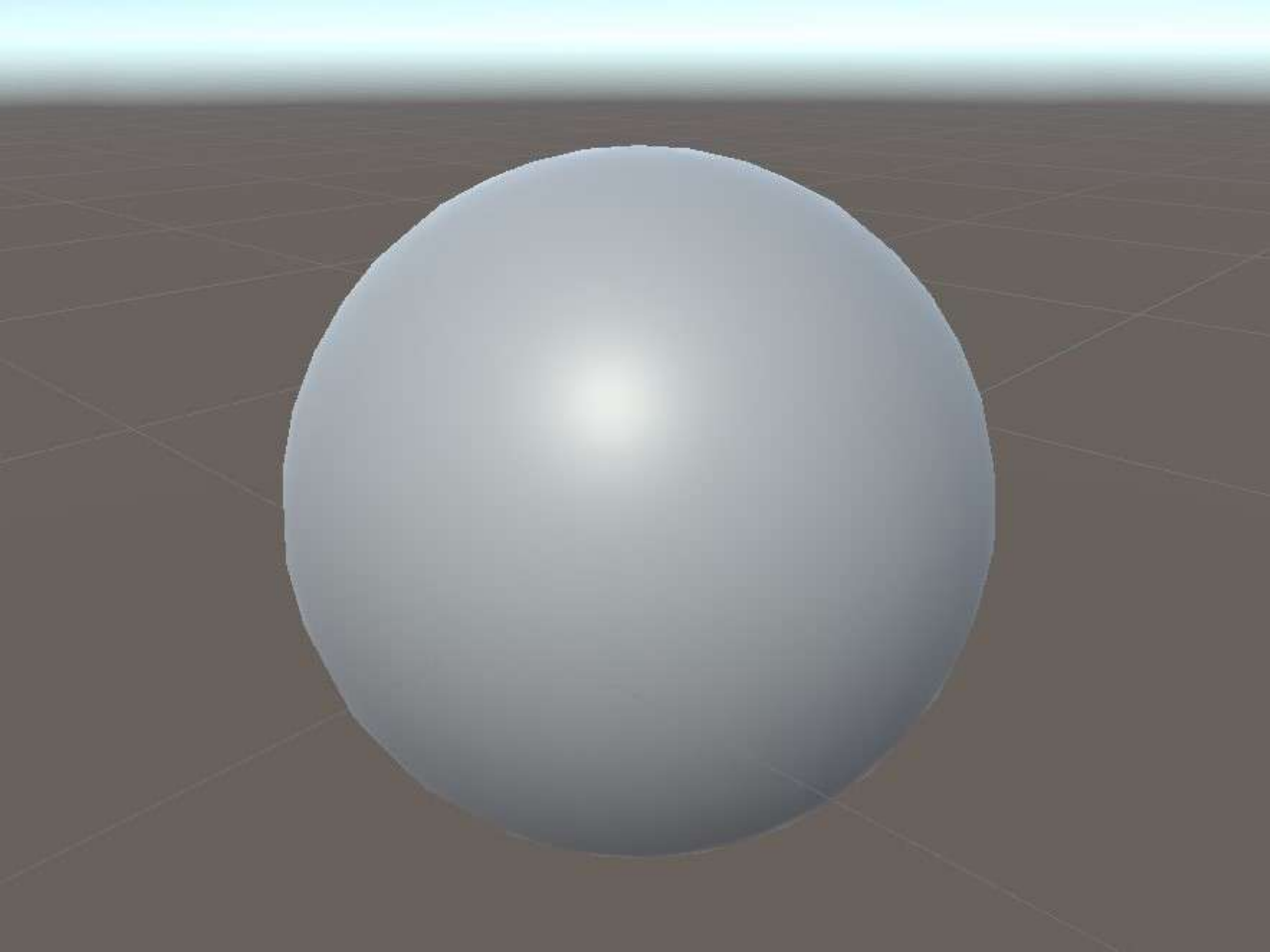}
    \includegraphics[height=0.94in]{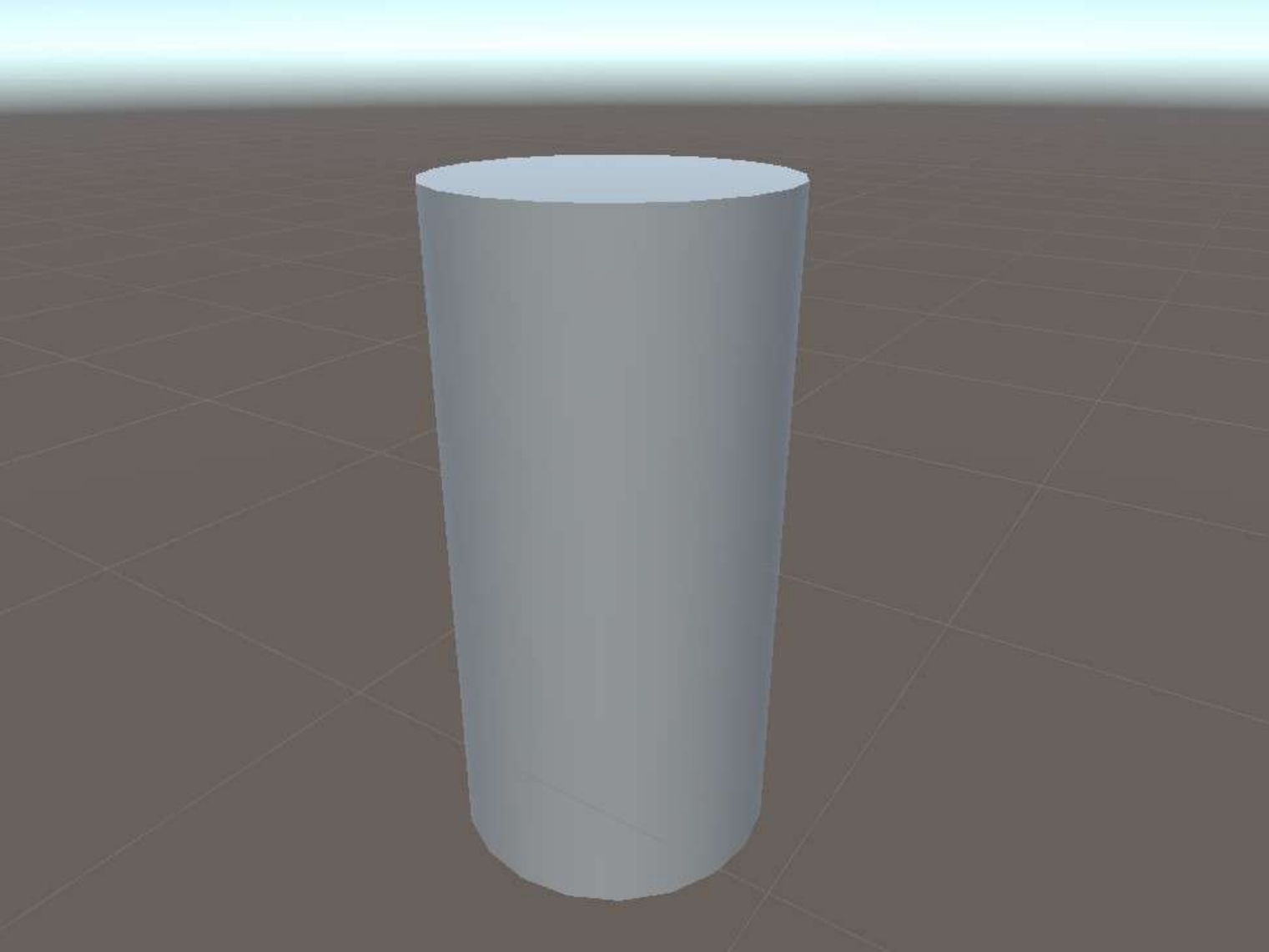}
    \includegraphics[height=0.94in]{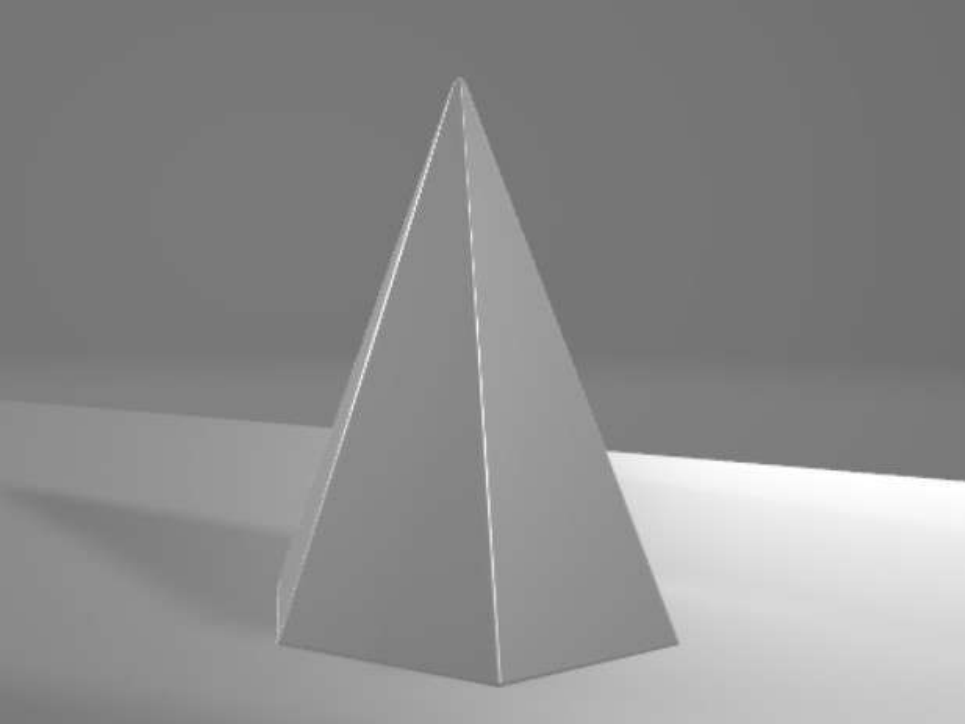}
    \includegraphics[height=0.94in]{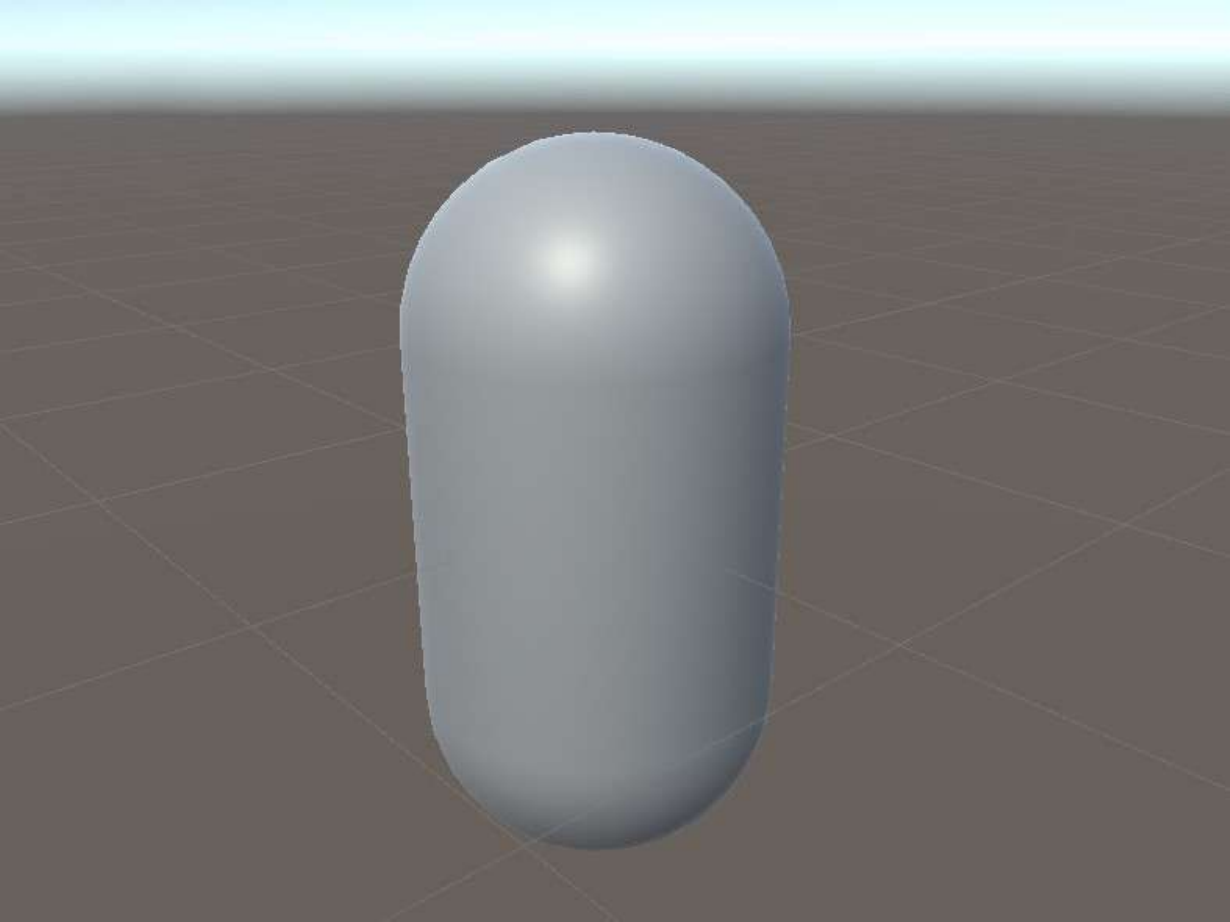}
    \includegraphics[height=0.94in]{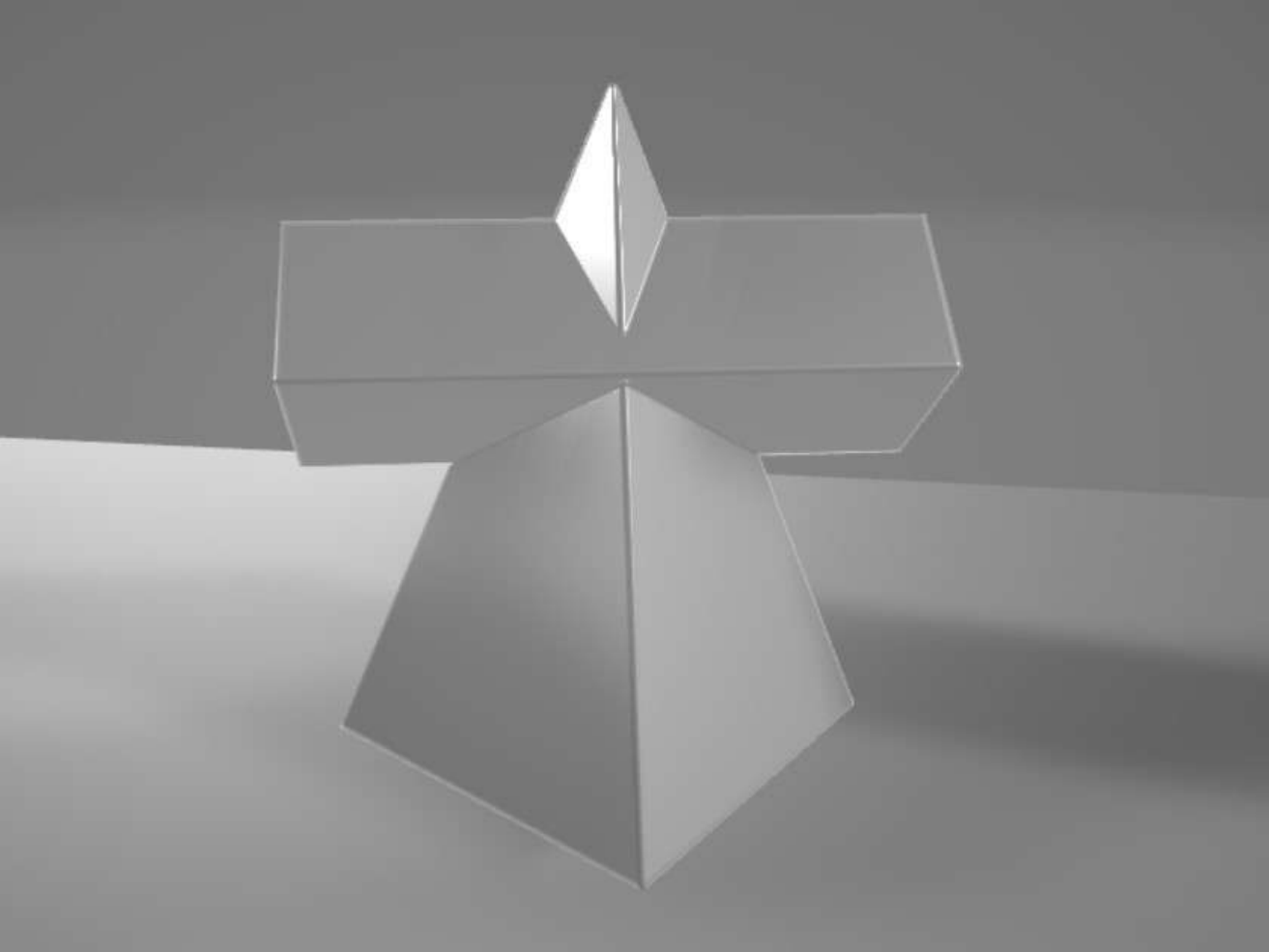}
    \includegraphics[height=0.94in]{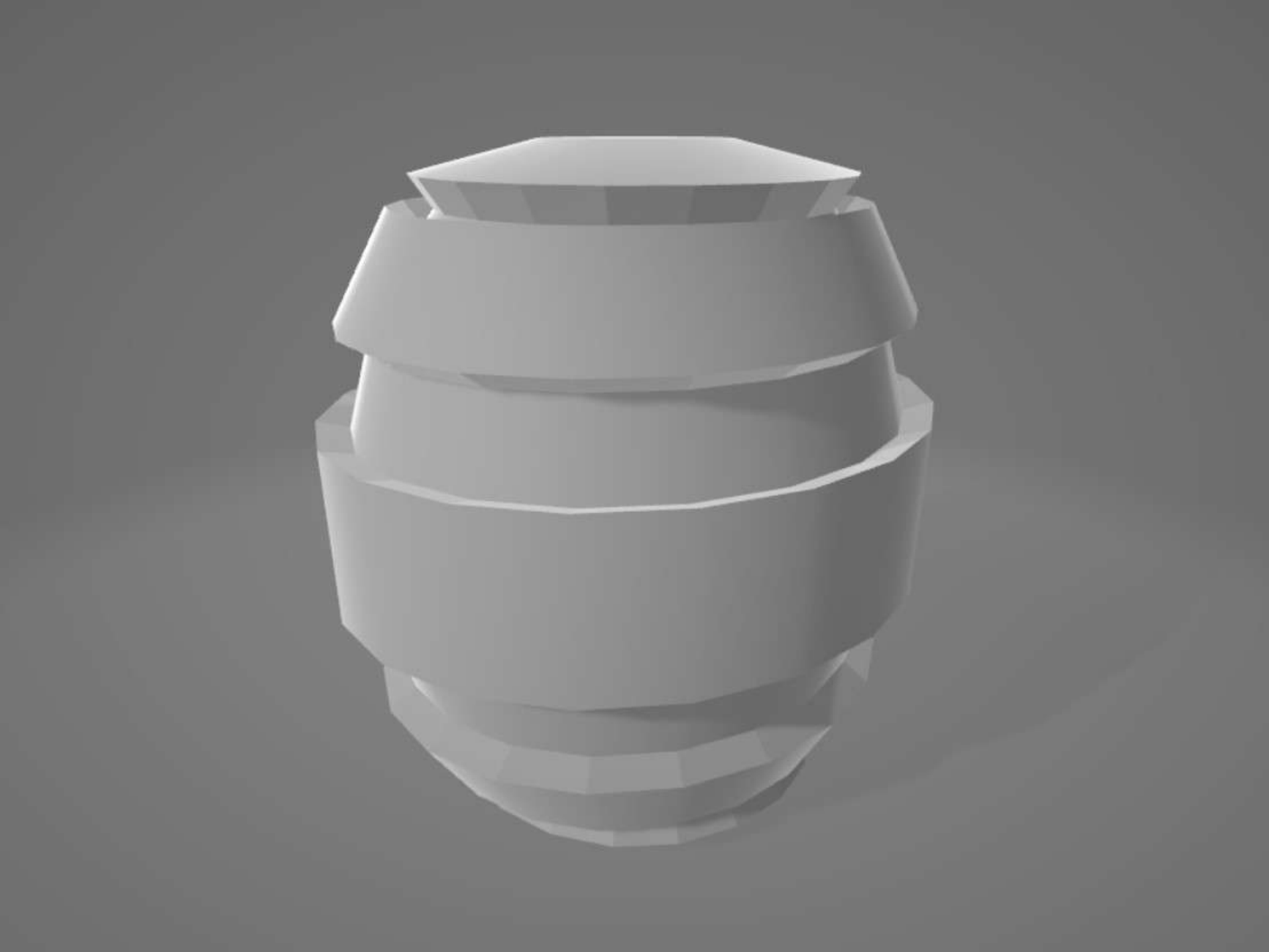}
    \includegraphics[height=0.94in]{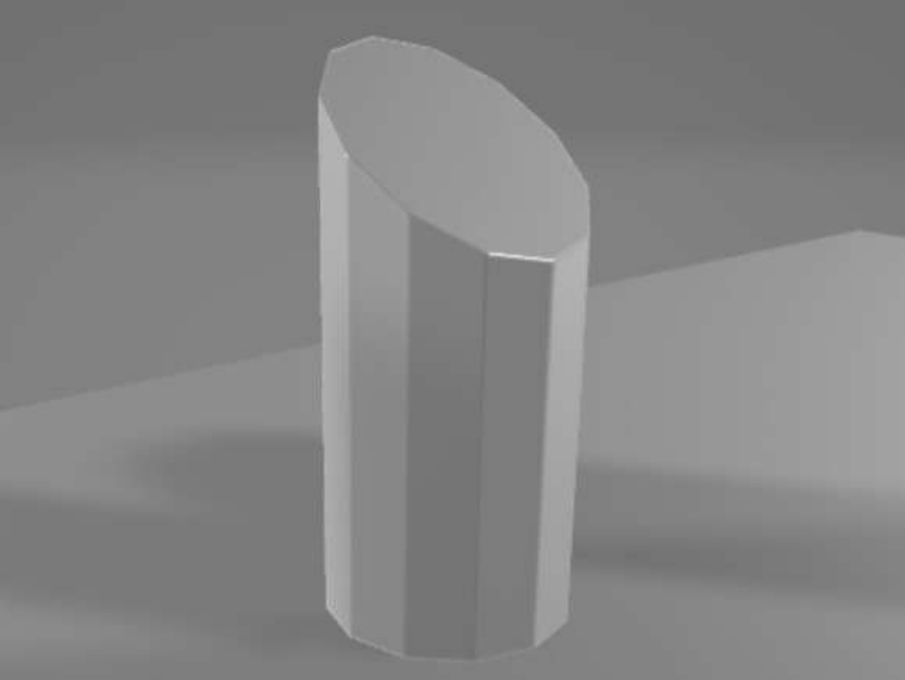}
    \includegraphics[height=0.94in]{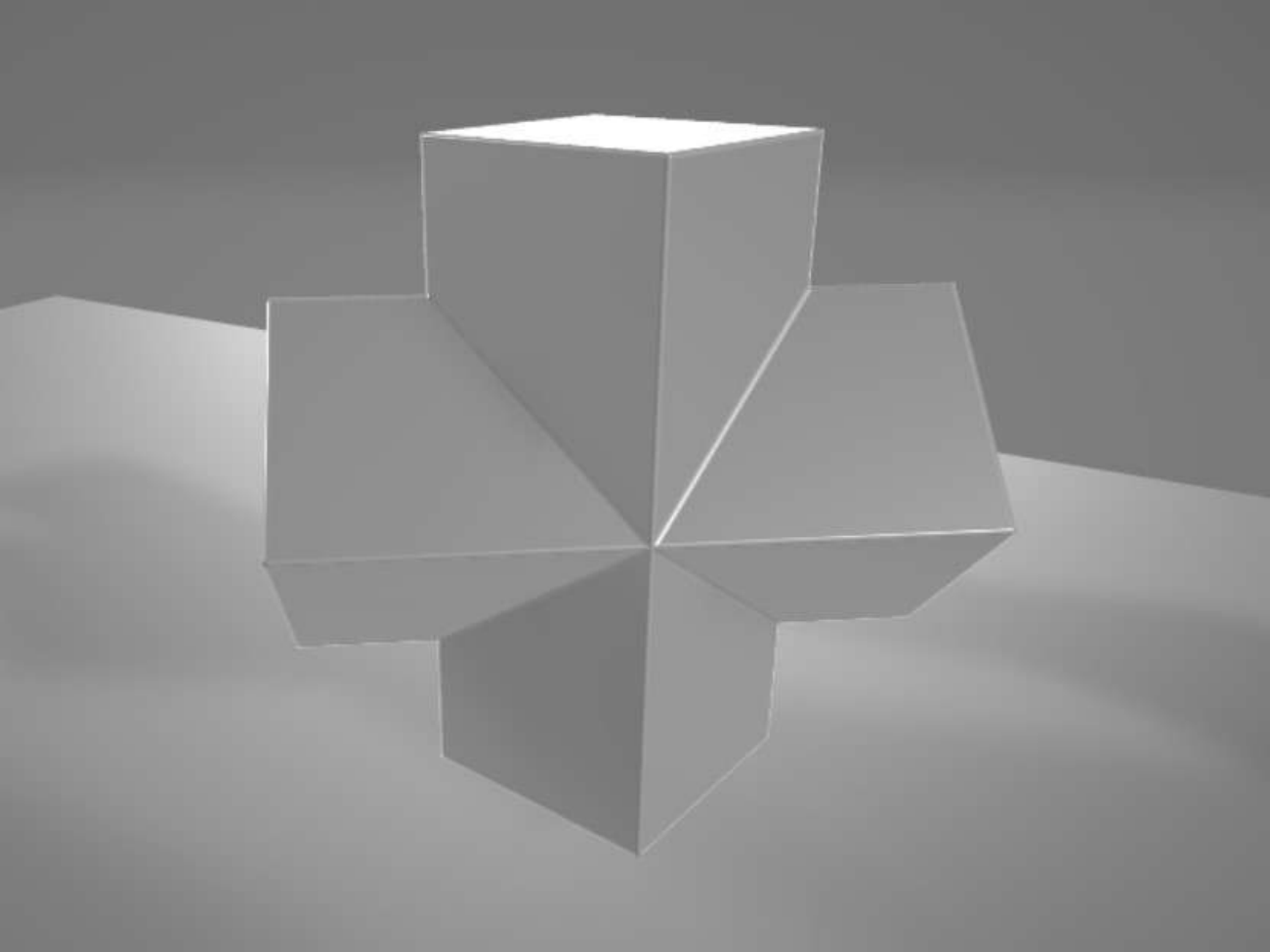}
    \includegraphics[height=0.94in]{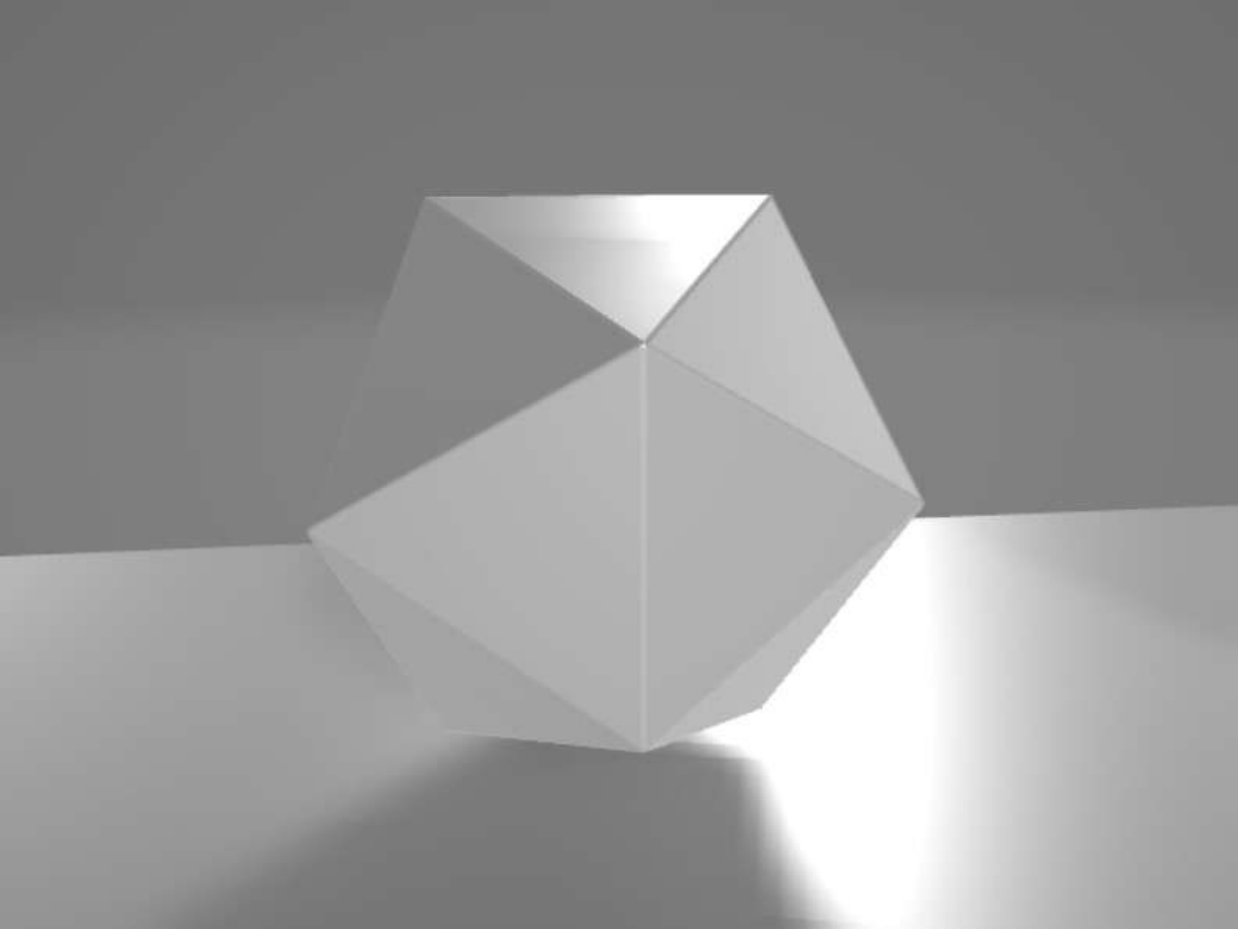}
    \includegraphics[height=0.94in]{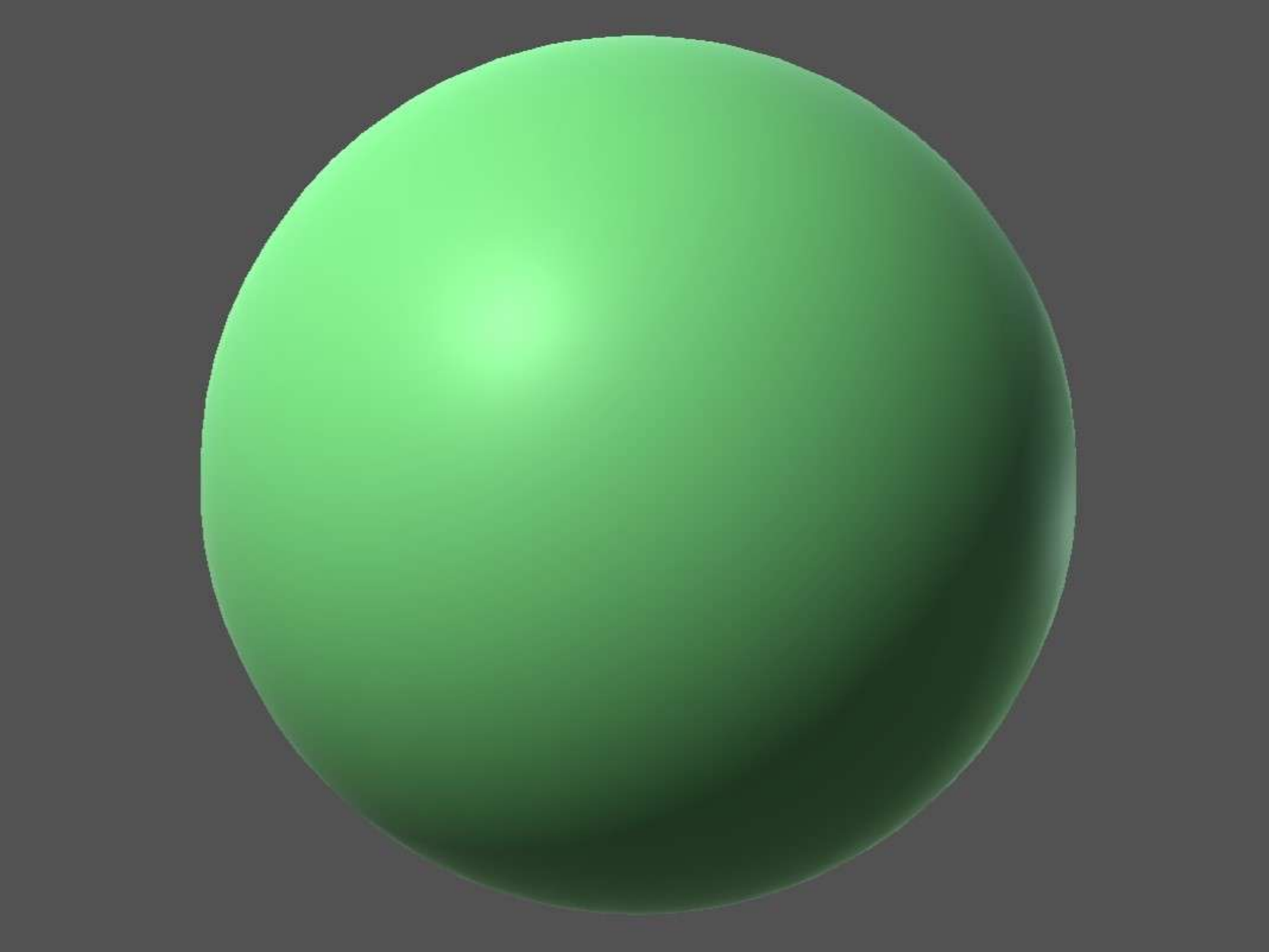}
    \includegraphics[height=0.94in]{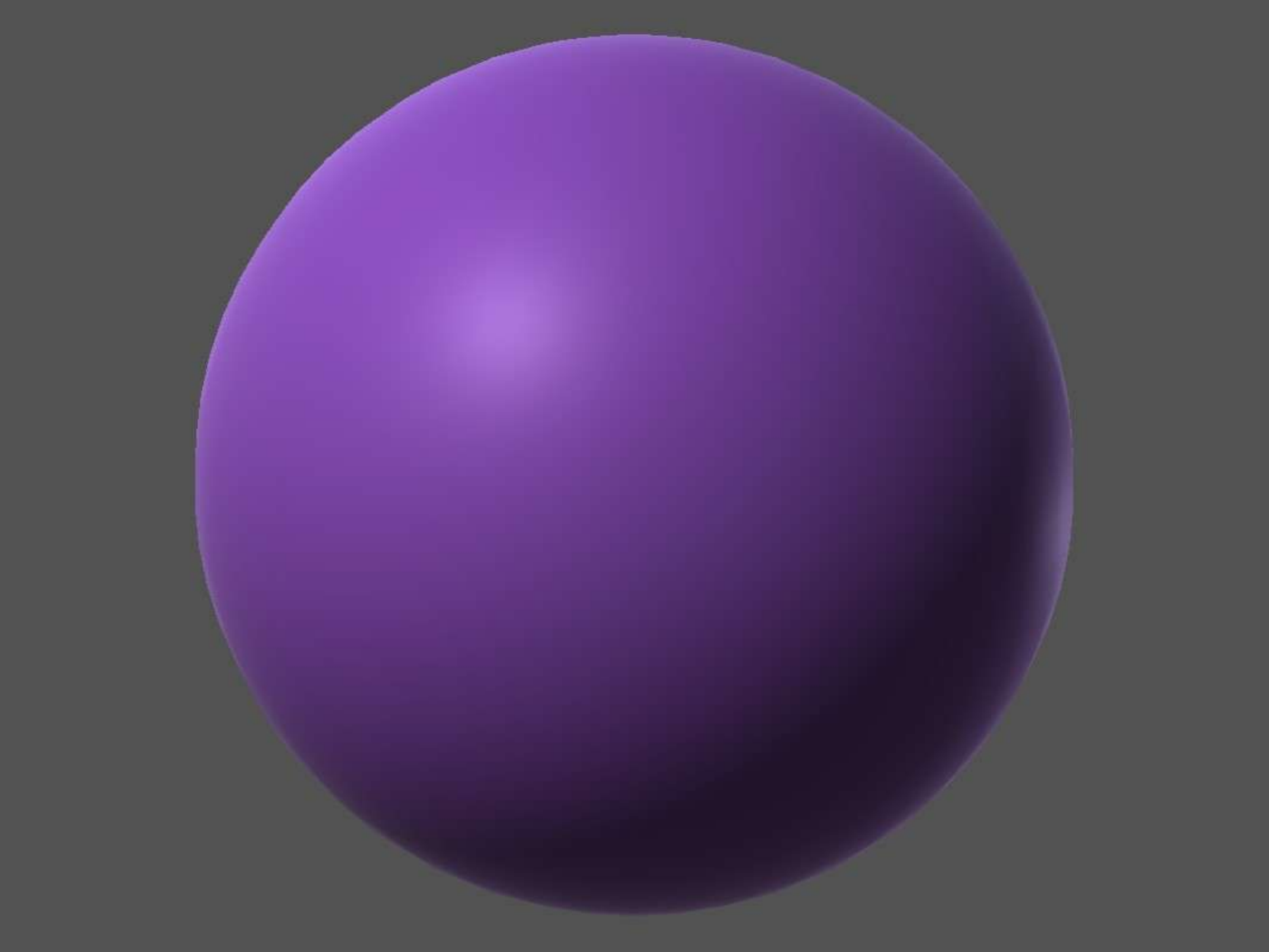}
    \includegraphics[height=0.94in]{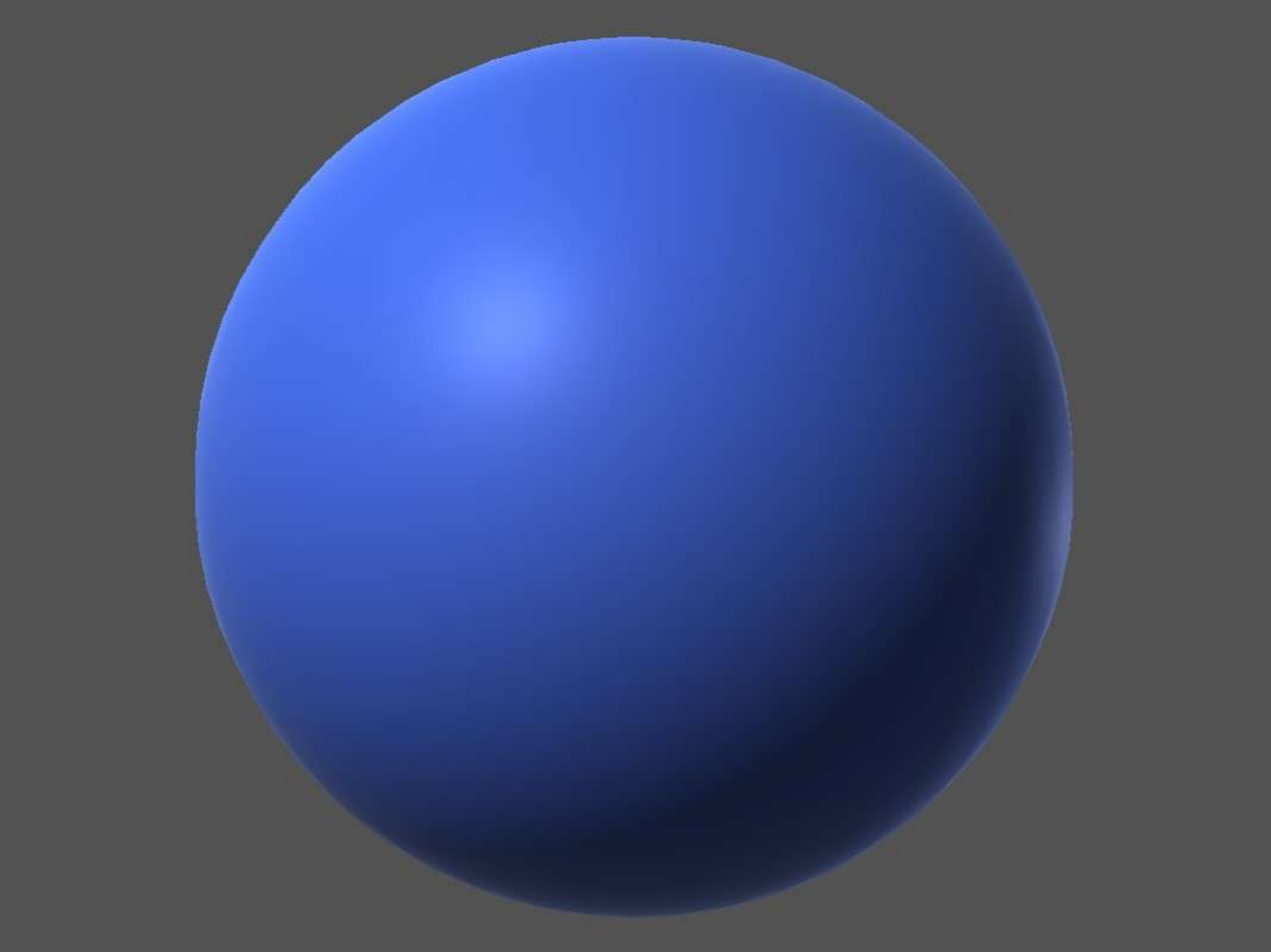}
    \includegraphics[height=0.94in]{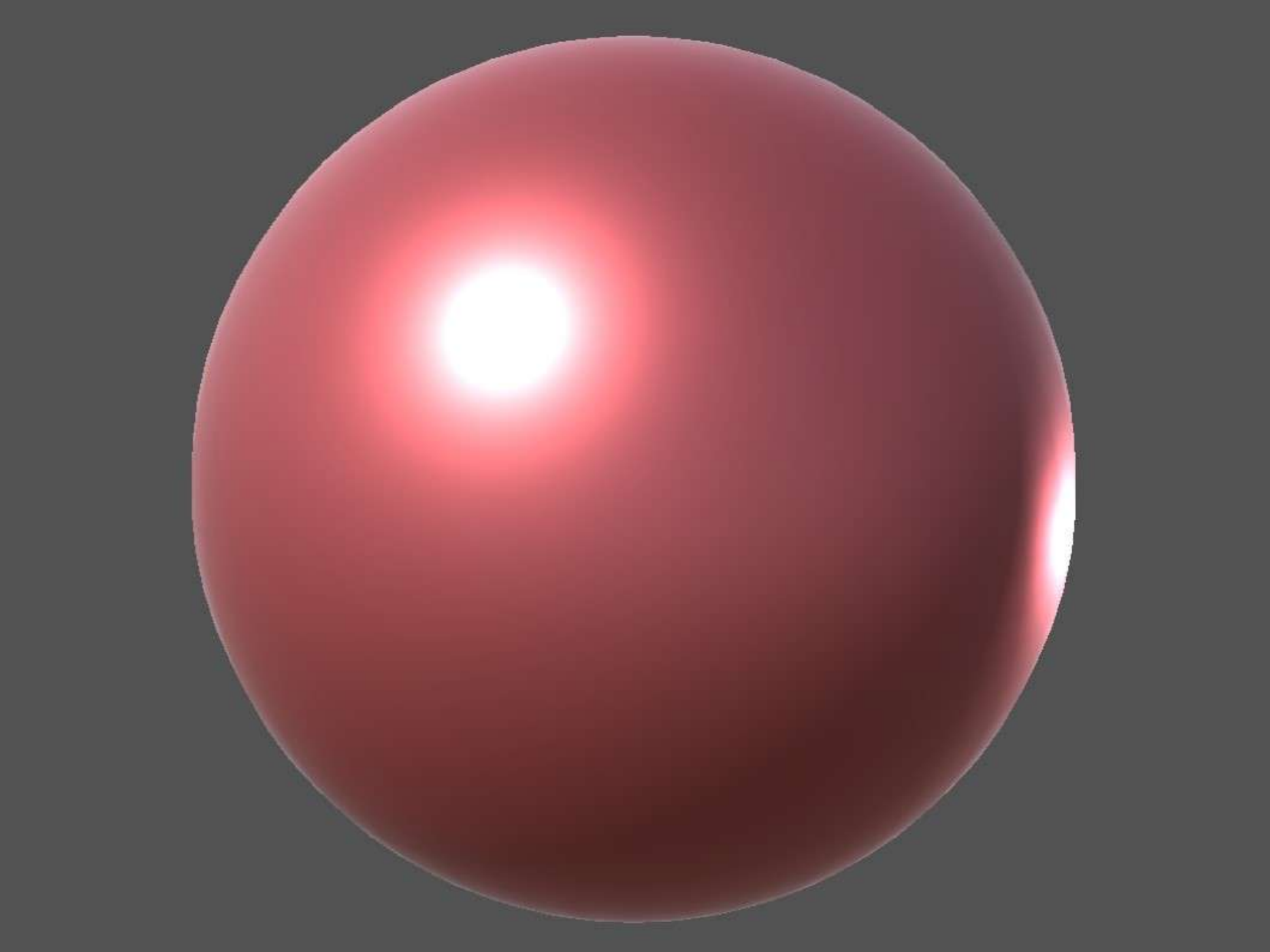}
    \includegraphics[height=0.94in]{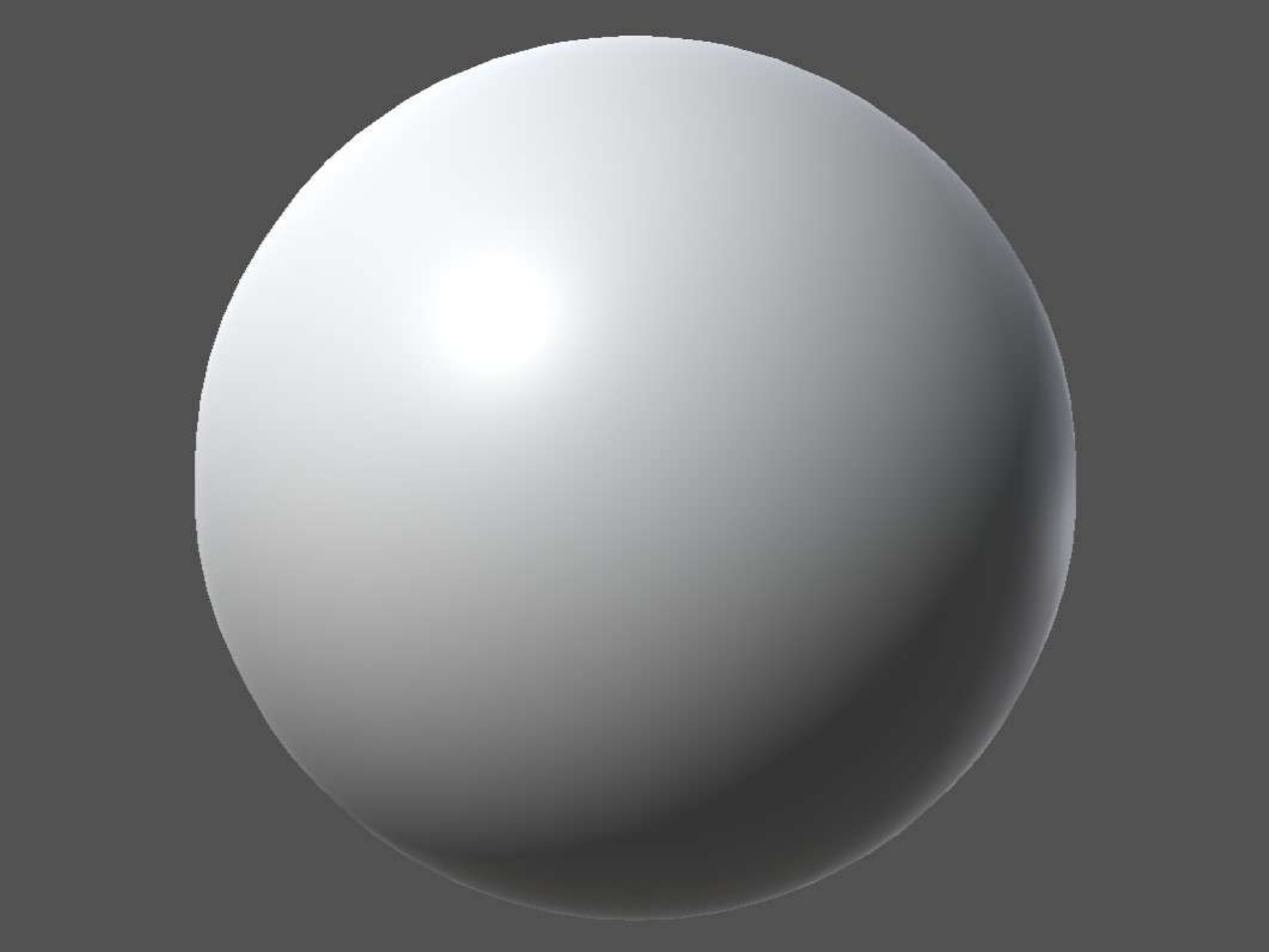}
    \includegraphics[height=0.94in]{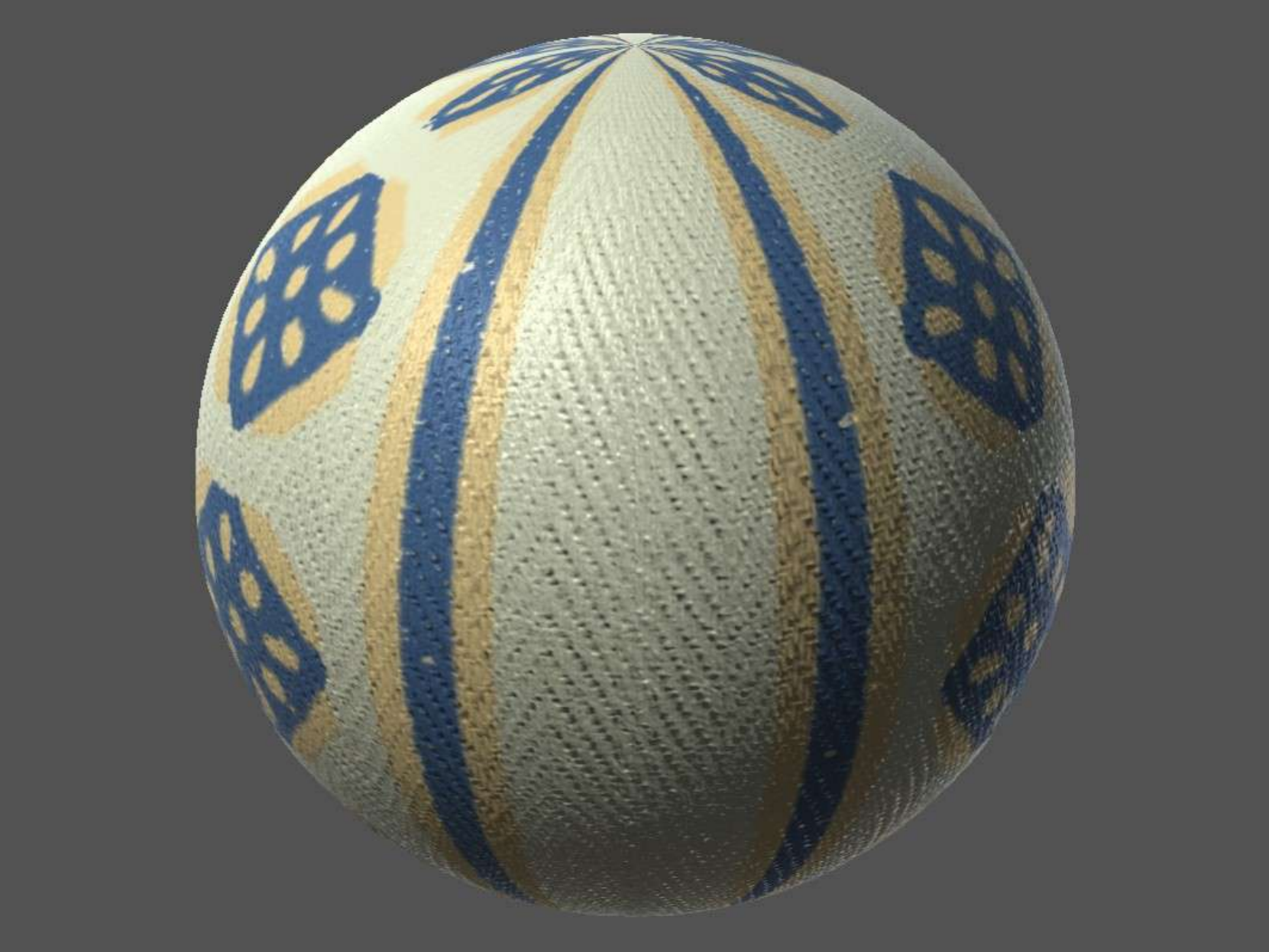}
    \includegraphics[height=0.94in]{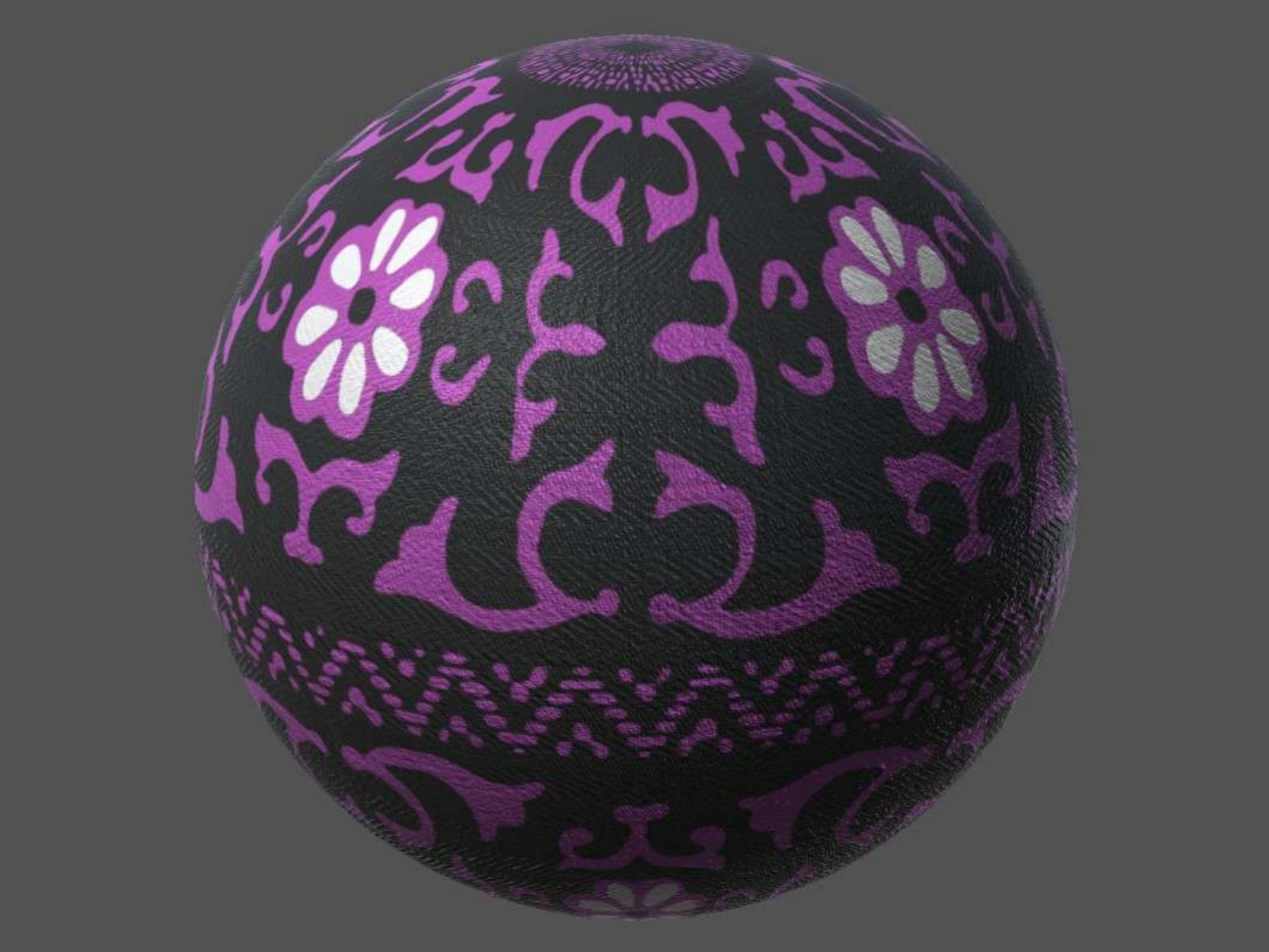}
    \includegraphics[height=0.94in]{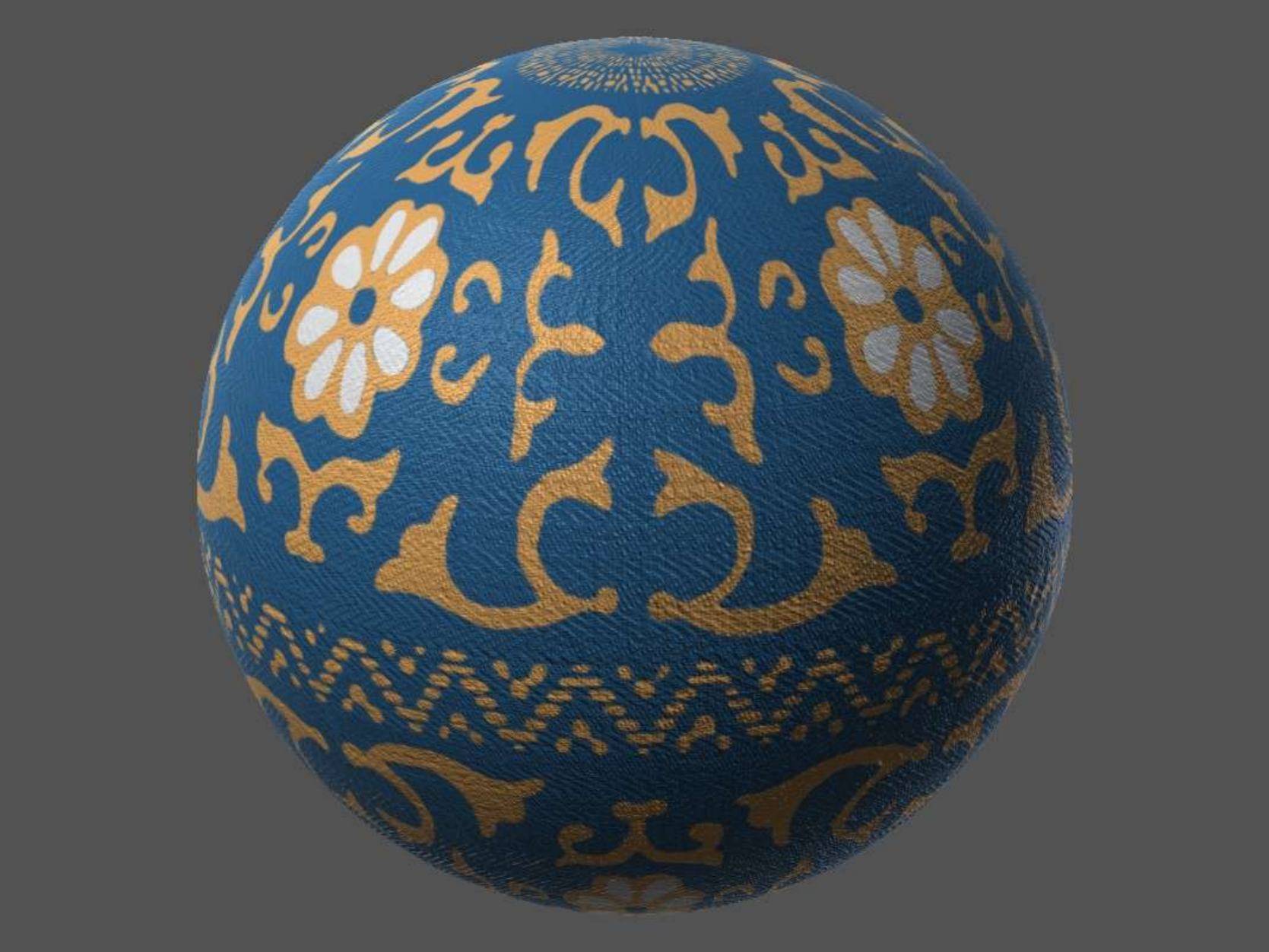}
    \includegraphics[height=0.94in]{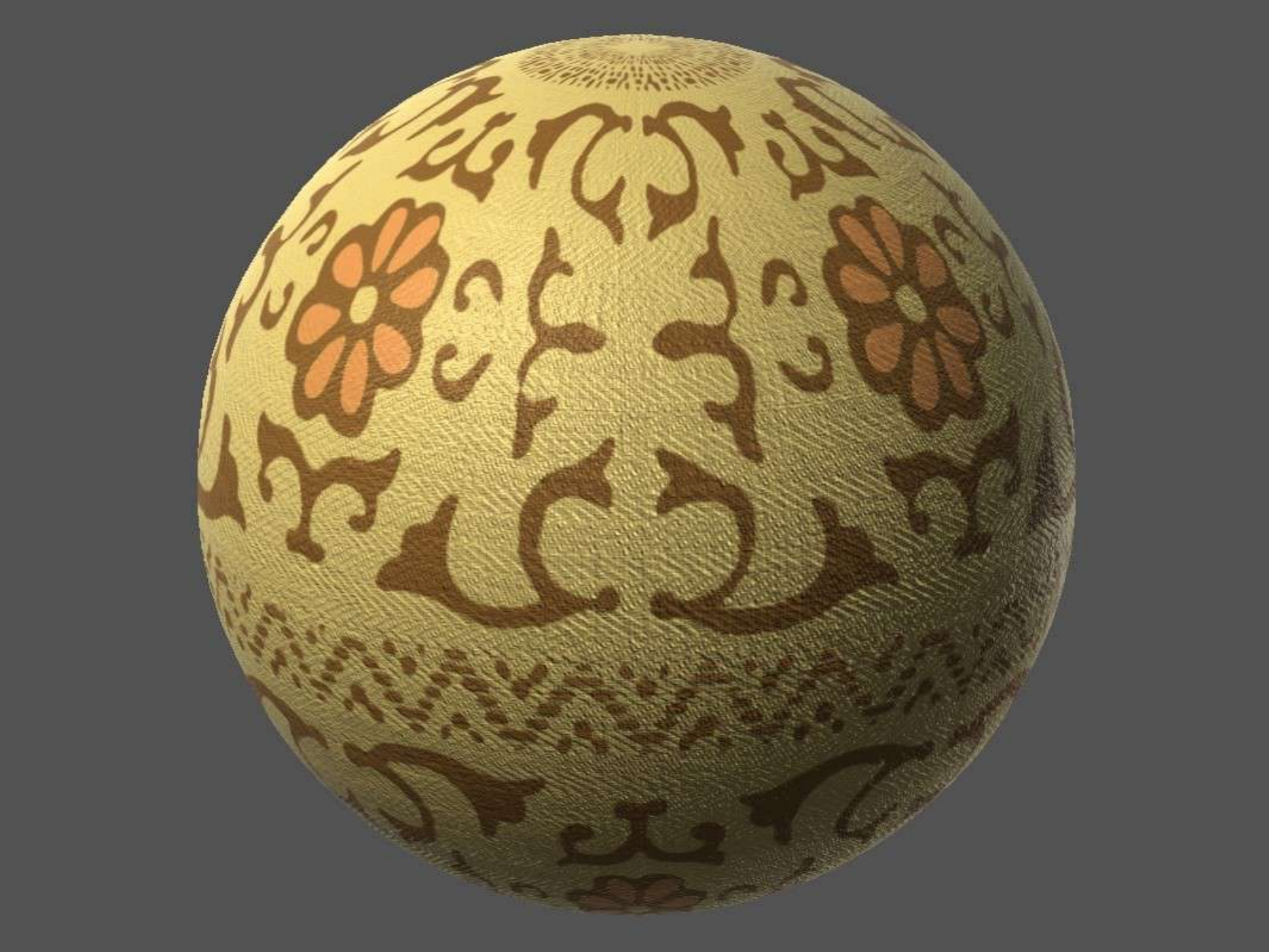}
    \includegraphics[height=0.94in]{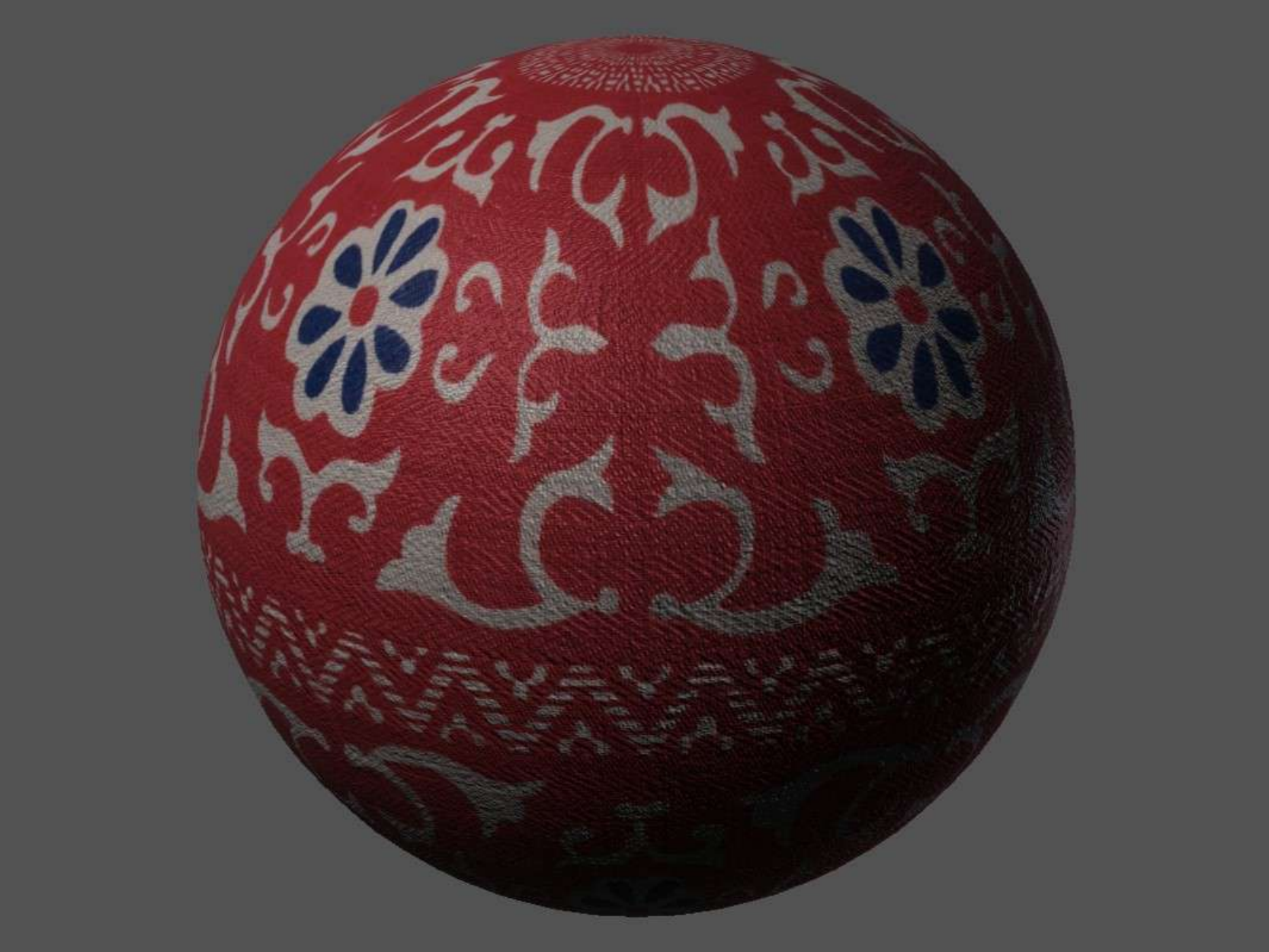}
    \caption{Images of 10 object shapes and 10 object colors presented in the pre-experiment survey. The first and third row include shapes and colors which we expected were easily identifiable. Shapes and colors in the second and fourth row were expected to be more difficult to reference verbally. Based on the survey results, shapes and colors from the last column were not included. The names of the remaining shapes and colors are listed as follows. First row: cube, sphere, cylinder, pyramid; Second row: pyramid cuboid, barrel, truncated cylinder, cross; Third row: green, purple, blue, red; Fourth row: white pattern, purple pattern, blue pattern, yellow pattern.}
    \label{fig:shapes-and-colors}
}

\maketitle

\vfill


\appendix 


\section{Pre-Experiment Survey and Findings}
\label{appendix:pre-experiment}

Prior to the user study introduced in this paper, we distributed an online pre-experiment survey to determine which object shapes and object colors were easy to reference verbally and which were not. 

\subsection{Participants}

Of the 21 participants who responded to this survey, 52.6\% were male and 47.4\% were female. The age of participants ranged from 19 to 43, and no disability of any form was reported.

\subsection{Procedure}

In the survey, participants were given images of ten object shapes and images of ten object colors, among which we expected five shapes and five colors to be easily identifiable, and the rest to be more difficult to reference verbally. For each object shape or color, they were asked to name it in no more than two words. The ten images within each section are presented to participants in random order. If participants found some of the object shapes or colors hard to describe, they had the option to choose not to describe it. Subsequently, participants were given ten images of different target objects among a set of distractor objects in the background. The ten target objects were chosen such that all object shapes and all object colors were included, and participants were asked to name the target object to distinguish it from other objects in the scene in no more than three words. Finally, participants were asked to provide their name and preferred means of communication if they wish so that they could be notified of the subsequent object selection user study.

\subsection{Results}

\Cref{fig:shapes-and-colors} shows the images of all object shapes and colors that were included in the survey. According to the survey results, 100\%, 90.5\%, 85.8\%, 76.2\%, 52.4\% of participants referred to objects in the first row as `cube', `sphere', `cylinder', `pyramid', and `capsule' respectively. Based on these results, we dropped the capsule object shape and included only the first four objects as easily identifiable objects in the user study. Similarly, 85.7\%, 81\%, 71.4\%, 57.1\%, and 47.6\% were unable to reference object shapes in the second row verbally. We dropped the last shape and kept the first four shapes in the second row as object shapes which were difficult to reference verbally. In the survey, 9.6\% of participants referred to the first object as `pyramid', 9.6\% referred to the second object as `barrel', 4.8\% referred to the third object as `truncated cylinder', 19.2\% referred to the fourth object as `cross'. Therefore, we assigned the names `pyramid cuboid', `barrel', `truncated cylinder', and `cross' to the first four shapes in the second row in the user study.

For object colors, 100\%, 100\%, 95.5\%, 76.1\%, 52.6\% of participants referred to the colors in the third row of \Cref{fig:shapes-and-colors} as `green', `purple', `blue', `red', and `white', and only the first four colors were included in the study. 90.5\%, 81\%, 76.2\%, 71.4\%, and 66.7\% of participants were not able to reference the colors on the last row verbally, and the last color was excluded from the study. For participants who attempted to reference the colors, as the names chosen for each color were drastically different, we decided to name them as `white pattern', `purple pattern', `blue pattern', and `yellow pattern' to maintain consistency.

In the third section of the survey, participants were asked to name objects with a certain shape and color to distinguish with other objects. For some objects, both the shape and color was easy to reference, examples of which included the purple cube (90.4\% participants were able to describe it as `purple' and 95.2\% were able to describe it as `cube'), the white pyramid (66.8\% described it `white' and 81.2\% described it as a `pyramid'), and the blue capsule (71.4\% described it `blue' and 57.2\% described it as a `capsule'). Some objects had either an unfamiliar shape or an unfamiliar color, for example the white pattern sphere (only 4.8\% participants described `white pattern' and 67.1\% described it as a `sphere'), the red pattern cylinder (4.8\% participants described `red pattern' and 62.4\% described it as a `cylinder'), the green cross (57.5\% participants described `green' and 28.7\% described it as a `cross'), and the red pyramid cuboid (28.8\% participants described `red' and only 4.8\% described `pyramid cuboid'). Other objects had both an unfamiliar shape and an unfamiliar color, such as the blue pattern deltahedron (none of the participants described `blue pattern' and only 9.6\% described `deltahedron'), the yellow pattern truncated cylinder (only 4.8\% participants described `yellow pattern' and 4.8\% described `truncated cylinder'), and the purple pattern barrel (none of the participants described `purple pattern' and only 9.6\% described `barrel').


\subsection{Implications}

The pre-experiment survey provided evidence to determine which object shapes and colors were easy to reference verbally and which were not. Based on the 4 shapes and 4 colors which were easy to reference, we constructed the \textsc{Low} perplexity condition; Based on the 4 shapes and colors which were difficult to reference, we constructed the \textsc{High} perplexity condition; Based on the 2 shapes and colors which were most easy to reference, and the 2 shapes and colors which were most difficult to reference, we constructed the \textsc{Medium} perplexity condition. For further reference, the complete version of the pre-experiment survey and anonymized results can be found on OSF in the `Supplemental Materials' folder within the linked GitHub directory.


